# On the Capacity of Free-Space Optical Intensity Channels

Amos Lapidoth       Stefan M. Moser       Michèle A. Wigger*

November 3, 2018

## Abstract

New upper and lower bounds are presented on the capacity of the free-space optical intensity channel. This channel is characterized by inputs that are nonnegative (representing the transmitted optical intensity) and by outputs that are corrupted by additive white Gaussian noise (because in free space the disturbances arise from many independent sources). Due to battery and safety reasons the inputs are simultaneously constrained in both their average and peak power. For a fixed ratio of the average power to the peak power the difference between the upper and the lower bounds tends to zero as the average power tends to infinity, and the ratio of the upper and lower bounds tends to one as the average power tends to zero.

The case where only an average-power constraint is imposed on the input is treated separately. In this case, the difference of the upper and lower bound tends to 0 as the average power tends to infinity, and their ratio tends to a constant as the power tends to zero.

## 1 Introduction

We consider a channel model for short-range optical communication in free space such as the infrared communication between electronic handheld devices. We assume a channel model based on *intensity modulation*, where the input signal modulates the optical intensity of the emitted light. Thus, the input signal is proportional to the light intensity and is therefore nonnegative. We further assume that at the receiver a front-end photodetector measures the incident optical intensity of the incoming light and produces an output signal which is proportional to the detected intensity. We model the ambient light conditions by a Gaussian disturbance. Moreover, we assume that the line-of-sight component is dominant and ignore any effects due to multiple-path propagation like fading or inter-symbol interference.[1]

Optical communication is restricted not only by battery power but also, for safety reasons, by the maximum allowed peak power. We therefore consider simultaneously two constraints: an average-power constraint $\mathcal{E}$ and a maximum allowed peak power A. The situation where only a peak-power constraint is imposed, corresponds to $\mathcal{E} = \mathrm{A}$. The case of only an average-power constraint is treated separately.

---

*A. Lapidoth and M. A. Wigger are with the Department of Information Technology and Electrical Engineering, Swiss Federal Institute of Technology (ETH) in Zurich, Switzerland; S. M. Moser is with the Department of Communication Engineering, National Chiao Tung University (NCTU) in Hsinchu, Taiwan. The work of S. M. Moser was supported in part by ETH under TH-23 02-2 and in part by the National Science Council, Taiwan, under NSC 96-2221-E-009-012-MY3.
Presented in part at the Winterschool on Coding and Information Theory 2005, Bratislava, Slovakia.

[1]For more details on the channel model see Section 2.

The described system is called the *free-space optical intensity channel* and has previously been studied in [1], [2], [3], [4], [5]. In [3] it has been proved that the capacity-achieving probability measure for this channel is discrete, and in [4], [5] upper and lower bounds on this channel's capacity have been derived. Related channel models used to describe optical communication are the Poisson channel, see [6], [7], [8], [2] for the discrete-time channel and [9], [10], [11], [12], [13], [14], [15] for the continuous-time channel, and a variation of the free-space optical optical intensity where the noise depends on the input [2, Chapter 4], [16].

In this work we present new upper and lower bounds on the capacity of the free-space optical intensity channel and study the capacity's asymptotic behavior at high and low powers. The maximum gap between the upper and lower bounds never exceeds 1 nat when the ratio of the average-power constraint to the peak-power constraint is larger than 0.03 or when only an average-power constraint is imposed. For the case of average-power and peak-power constraints, asymptotically when the available average and peak power tend to infinity with their ratio held fixed, the upper and lower bound coincide, *i.e.*, their difference tends to 0. When the available average and peak power tend to 0, with their ratio held fixed, the ratio of the upper and lower bound tends to 1. For the case of only an average-power constraint the proposed upper and lower bound coincide asymptotically for high power, *i.e.*, their difference tends to 0 as the power tends to infinity. At low power their ratio tends to $2\sqrt{2}$.

The derivation of the upper bounds is based on a general technique introduced in [17] using a dual expression of mutual information. We will not state it in its full generality but only in the form needed in this paper. For more details and for a proof see [17, Sec. V], [2, Ch. 2].

**Proposition 1.** *Assume a memoryless channel with input alphabet $\mathcal{X} = \mathbb{R}_0^+$ and output alphabet $\mathcal{Y} = \mathbb{R}$ where conditional on the input $x \in \mathcal{X}$ the distribution on the output $Y$ is denoted by the probability measure $W(\cdot|x)$.[2] Then, for arbitrary distribution $R(\cdot)$ over $\mathcal{Y}$, the channel capacity under a peak-power constraint $\mathrm{A}$ and an average-power constraint $\mathcal{E}$ is upper-bounded by*

$$\mathrm{C}(\mathrm{A}, \mathcal{E}) \leq \sup_Q \mathsf{E}_Q\big[D\big(W(\cdot|X)\big\|R(\cdot)\big)\big], \tag{1}$$

*where the supremum is taken over all probability laws $Q$ on the input $X$ satisfying $Q(X > \mathrm{A}) = 0$ and $\mathsf{E}_Q[X] \leq \mathcal{E}$. Here, $D(\cdot\|\cdot)$ stands for the relative entropy [18, Ch. 2].*

*Proof.* See [17, Sec. V]. □

There are two challenges in using (1). The first is in finding a clever choice of the law $R$ that will lead to a good upper bound. The second is in upper-bounding the supremum on the right-hand side of (1). To handle this second challenge we shall resort to some further bounding, *e.g.*, Jensen's inequality [18, Ch. 2.6].

---

[2]The proposition requires certain measurability assumptions on the law $W(\cdot|\cdot)$ which we omit for simplicity. However, the channel law under consideration (see its density (9) ahead) satisfies these assumptions. See [17, Sec. V], [2, Ch. 2] for a description of the assumptions.

To derive the lower bounds we apply two different techniques: one for the high-power regime, and the other for the low-power regime. For high powers we use the entropy power inequality (see Lemma 16) and the theory of entropy maximizing distributions [18, Ch. 11]. Asymptotically, the differences of these lower bounds and some of the upper bounds derived using duality tend to 0 as the power tends to infinity, and thus the bounds are tight at high power. At low powers we lower-bound capacity considering binary input distributions; a choice which was inspired by [19] and [3]. In the cases involving a peak-power constraint, the asymptotic behavior of the corresponding mutual information is studied using [20]. When only an average-power constraint is imposed, a lower bound on the asymptotic behavior of the mutual information is derived. In the cases involving a peak-power constraint the asymptotic expression of the mutual information for binary inputs and some of the duality-based upper bounds are asymptotically tight at low power, *i.e.*, their ratio tends to one as the power tends to 0. When only an average-power constraint is imposed, the derived lower bound on the asymptotic expression of the mutual information is not tight with any of the duality-based upper bounds. Indeed, the ratio of the best upper bound with this lower bound tends to $2\sqrt{2}$ when the average power tends to 0.

The rest of the paper is structured as follows. After some remarks about notation at the end of this section, we define the considered channel model in detail in the subsequent Section 2. Section 3 contains some mathematical preliminaries. In Section 4 we state our main results, *i.e.*, the upper and lower bounds on channel capacity and the asymptotic results. The detailed derivations of the lower bounds, the upper bounds, and the asymptotic results can be found in Sections 5, 6, and 7, respectively.

For random quantities we use uppercase letters and for their realizations lower-case letters. Scalars are typically denoted using Greek letters or lower-case Roman letters. A few exceptions are the following symbols: C stands for capacity, $D(\cdot\|\cdot)$ denotes the relative entropy between two probability measures, and $I(\cdot;\cdot)$ stands for the mutual information. Moreover, the capitals $Q$, $W$, and $R$ denote probability measures:

- $Q(\cdot)$ denotes a generic probability measure on the channel input;
- for any input $x \in \mathcal{X}$, $W(\cdot|x)$ represents a probability measure on the channel output when the channel input is $x$;
- $R(\cdot)$ denotes a generic probability measure on the channel output.

The expression $I(Q, W)$ stands for the mutual information between input $X$ and output $Y$ of a channel with transition probability measure $W$ when the input has distribution $Q$, *i.e.*,

$$I(Q, W) \triangleq I(X; Y). \qquad (2)$$

The symbol $\mathcal{E}$ denotes average power and A stands for peak power. We denote the mean-$\eta$, variance-$\sigma^2$ real Gaussian distribution by $\mathcal{N}_\mathbb{R}(\eta, \sigma^2)$. All rates specified in this paper are in nats per channel-use, and all logarithms are natural logarithms.

# 2 Channel Model

## 2.1 Physical Description

In free-space optical communication the input signal usually is transmitted by means of light emitting diodes (LED) or laser diodes (LD). Conventional and most inexpensive diodes emit infrared light of wavelength between 850 and 950 nanometers. For such high frequencies, practical systems often apply intensity modulation where the transmitter modulates the optical intensity of the emitted light, and hence the input signal is proportional to the optical intensity. The receiver first measures the incident optical intensity of the incoming light by means of a front-end photodetector and produces an output signal which is proportional to the detected intensity. Based on this output signal the receiver decodes the transmitted data.

For our model we neglect the impact of fading or inter-symbol interference due to multiple-path propagation and assume that the direct line-of-sight path is dominant. In the absence of a protective medium like, *e.g.*, a fiber cable, the dominant noise source is assumed to be strong ambient light. Even if optical filters are applied to reduce the impact of this noise, it typically has much larger power than the actual signal and causes high-intensity shot noise in the output signal. In a first approximation this shot noise can be assumed to be additive and independent of the signal itself.

The maximum allowed optical peak power of the transmitted signal has to be constrained, *e.g.*, to guarantee eye safety. Moreover, to increase battery life-time, we also constrain the allowed optical average power. Since the optical power is proportional to the optical intensity, and the input signal modulates the optical intensity, in the described system the optical power is proportional to the input signal. Thus, the constraints on the optical peak and average power have to be transformed into peak and average constraints on the input signal (and not on its square as usual in radio communication). For a more detailed description of the free-space optical intensity channel see [21].

## 2.2 Mathematical Channel Model

We will now translate the above physical channel description into a simplified time-discrete channel model. The time-$k$ channel output $\tilde{Y}_k$ is given by

$$\tilde{Y}_k = x_k + \tilde{Z}_k. \tag{3}$$

Here, $x_k$ denotes the time-$k$ channel input and the random process $\{\tilde{Z}_k\}$ models the additive noise. As described above this noise is mainly caused by strong ambient light. We therefore approximate it as a constant intensity term $\eta$ and some intensity-fluctuations around $\eta$. Because these fluctuations are caused by many independent sources, we assume that they are independent and identically distributed (IID) zero-mean Gaussian with a given variance $\sigma^2 > 0$. *I.e.*,

$$\{\tilde{Z}_k\} \sim \text{IID } \mathcal{N}_{\mathbb{R}}(\eta, \sigma^2). \tag{4}$$

Since $\eta$ is constant, we may without loss of generality neglect it because the receiver can always subtract or add any constant signal. We then define a new channel output random variable

$$Y_k = x_k + Z_k, \tag{5}$$

where $\{Z_k\} \sim$ IID $\mathcal{N}_{\mathbb{R}}(0, \sigma^2)$. Notice that the new channel outputs $\{Y_k\}$ represent the fluctuations of the electrical output signal around its working point $\eta$.

Our channel model is memoryless and therefore we drop the time-index $k$. The channel output $Y$ is then given by

$$Y = x + Z, \tag{6}$$

where $x$ denotes the channel input that is proportional to the optical intensity and therefore cannot be negative,

$$x \in \mathbb{R}_0^+, \tag{7}$$

and where the additive noise is

$$Z \sim \mathcal{N}_{\mathbb{R}}(0, \sigma^2). \tag{8}$$

Hence, the conditional probability law $W(\cdot|x)$ of the output $Y$ given input $x \in \mathbb{R}_0^+$ has density

$$f_{Y|X}(y|x) = \frac{1}{\sqrt{2\pi\sigma^2}} e^{-\frac{(y-x)^2}{2\sigma^2}}, \qquad x \in \mathbb{R}_0^+, \ y \in \mathbb{R}. \tag{9}$$

It is important to note that, unlike the input, the output $Y$ may be negative since the noise introduced at the receiver can be negative.

The restrictions on the optical peak and average power are translated into a peak-power and an average-power constraint on the input, respectively:

$$\Pr[X > \mathrm{A}] = 0, \tag{10}$$
$$\mathsf{E}[X] \leq \mathcal{E}, \tag{11}$$

for some fixed parameters $\mathcal{E}, \mathrm{A} > 0$. Note that the average-power constraint is on the expectation of the channel input and not on its square. We denote the ratio between average power and peak power by $\alpha$,

$$\alpha \triangleq \frac{\mathcal{E}}{\mathrm{A}}, \tag{12}$$

where $0 < \alpha \leq 1$. Note that for $\alpha = 1$ the average-power constraint is inactive, in the sense that it has no influence on the capacity and is automatically satisfied. This means that $\alpha = 1$ corresponds to the case with only a peak-power constraint. Similarly, $\alpha \ll 1$ corresponds to a dominant average-power constraint and only a very weak peak-power constraint.

We denote the capacity of the described channel with peak-power constraint A and average-power constraint $\mathcal{E}$ by $\mathrm{C}(\mathrm{A}, \mathcal{E})$. The capacity is given by [22]

$$\mathrm{C}(\mathrm{A}, \mathcal{E}) = \sup_Q I(Q, W) \tag{13}$$

where the supremum is over all laws $Q$ on $X \geq 0$ satisfying $Q(X > \mathrm{A}) = 0$ and $\mathsf{E}_Q[X] \leq \mathcal{E}$.

When only an average-power constraint is imposed, capacity is denoted by $\mathrm{C}(\mathcal{E})$. It is given as in (13) except that the supremum is taken over all laws $Q$ on $X \geq 0$ satisfying $\mathsf{E}_Q[X] \leq \mathcal{E}$.

# 3 The $\mathcal{Q}$-Function

**Definition 2.** *The $\mathcal{Q}$-function is defined by*

$$\mathcal{Q}(\xi) \triangleq \int_{\xi}^{\infty} \frac{1}{\sqrt{2\pi}} e^{-\frac{t^2}{2}} \, dt, \qquad \forall \, \xi \in \mathbb{R}. \tag{14}$$

Some of the properties of this function are recalled in the following lemma.

**Lemma 3.** *The $\mathcal{Q}$-function satisfies*

$$\mathcal{Q}(-\xi) + \mathcal{Q}(\xi) = 1, \qquad \forall \, \xi \in \mathbb{R}, \tag{15}$$

*and*

$$\mathcal{Q}(0) = \frac{1}{2}, \tag{16}$$

*and is bounded by*

$$\frac{1}{\sqrt{2\pi}\xi} e^{-\frac{\xi^2}{2}} \left(1 - \frac{1}{\xi^2}\right) < \mathcal{Q}(\xi) < \frac{1}{\sqrt{2\pi}\xi} e^{-\frac{\xi^2}{2}}, \quad \xi > 0, \tag{17}$$

*and*

$$\mathcal{Q}(\xi) \leq \frac{1}{2} e^{-\frac{\xi^2}{2}}, \quad \xi \geq 0. \tag{18}$$

*Its first and second derivatives are given by*

$$\mathcal{Q}'(\xi) = -\frac{1}{\sqrt{2\pi}} e^{-\frac{\xi^2}{2}}, \qquad \forall \, \xi \in \mathbb{R}, \tag{19}$$

*and*

$$\mathcal{Q}''(\xi) = \frac{\xi}{\sqrt{2\pi}} e^{-\frac{\xi^2}{2}}, \qquad \forall \, \xi \in \mathbb{R}. \tag{20}$$

*Thus, $\mathcal{Q}(\cdot)$ is monotonically strictly decreasing for all $\xi \in \mathbb{R}$, strictly concave over $(-\infty, 0)$, and strictly convex over $(0, \infty)$.*

*Proof.* The proof of (15) and (16) follows because $\frac{1}{\sqrt{2\pi}} e^{-\frac{\xi^2}{2}}$ is symmetric around 0, and because it equals the density of a standard Gaussian random variable and hence integrates to 1. For a proof of the bounds (17) and (18) see [23, pp. 83–84]. □

**Lemma 4.** *Let $\xi_0, \gamma \geq 0$ be nonnegative constants, and let the function $f(\cdot)$ be defined as*

$$f(\xi) \triangleq 1 - \mathcal{Q}(\xi_0 + \xi) - \mathcal{Q}(\xi_0 + \gamma - \xi), \quad \xi \in [0, \gamma]. \tag{21}$$

*Then $f(\cdot)$ is strictly concave over $[0, \gamma]$ and symmetric around $\xi = \frac{\gamma}{2}$. Furthermore, it is increasing over $\left[0, \frac{\gamma}{2}\right]$, decreasing over $\left[\frac{\gamma}{2}, \gamma\right]$, and takes on its maximum value at $\xi = \frac{\gamma}{2}$.*

*Proof.* See Appendix A. □

**Lemma 5.** *For $\mu \geq \frac{1}{\sqrt{e}}$ and $\xi \geq 0$:*

$$\xi \mathcal{Q}(\xi - \mu) \leq \mu. \tag{22}$$

*Proof.* See Appendix B. □

**Lemma 6.** *For $\mu, \xi \geq 0$:*

$$1 - \mathcal{Q}(\xi - \mu) \leq 1 - \mathcal{Q}(-\mu) + \frac{\xi}{\mu} \mathcal{Q}(-\mu). \tag{23}$$

*Proof.* See Appendix C. □

# 4 Results

The results of this paper are partially based on the results in [24] and [2, Ch. 3].

**Lemma 7.** *Given peak-power constraint* A *and average-power constraint* $\mathcal{E}$ *the capacity* $C(A, \mathcal{E})$ *of the free-space optical intensity channel* (6) *has a unique input distribution* $Q^*$ *that achieves the supremum in* (13).

*Proof.* See [3]. □

Using this lemma together with the symmetry of the channel law ((6) and (8)) and the concavity of mutual information in the input distribution, the following lemma can be proved.

**Lemma 8.** *If the allowed average power* $\mathcal{E}$ *is larger than half the allowed peak power* A, *then the optimal input distribution* $Q^*$ *in* (13) *satisfies*

$$\mathsf{E}_{Q^*}[X] = \frac{1}{2}\mathrm{A}. \tag{24}$$

*Thus,*

$$\mathrm{C}(\mathrm{A}, \alpha \mathrm{A}) = \mathrm{C}\left(\mathrm{A}, \frac{\mathrm{A}}{2}\right), \quad \frac{1}{2} < \alpha \leq 1. \tag{25}$$

*Proof.* See Appendix D. □

To state our results we distinguish between three different cases:

- Case I: both an average-power and a peak-power constraint are imposed, with $\alpha \in \left(0, \frac{1}{2}\right)$;

- Case II: both an average- and a peak-power constraint are imposed, with $\alpha \in \left[\frac{1}{2}, 1\right]$;

- Case III: only an average-power constraint is imposed.

We present firm upper and lower bounds on the channel capacity in all three cases. In all three cases their gap tends to 0 as the available power tends to infinity, and thus, we can derive the asymptotic capacity at high power. We also present the asymptotics of capacity at low power. For case I and II we state them exactly, *i.e.*, we present asymptotic upper and lower bounds whose ratio tends to 1 as the power tends to 0. For case III we present asymptotic upper and lower bounds whose ratio tends to $2\sqrt{2}$ as the power tends to 0.

## 4.1 Bounds on Channel Capacity for Case I

**Theorem 9.** *If* $0 < \alpha < \frac{1}{2}$, *then* $C(A, \alpha A)$ *is lower-bounded by*

$$\mathrm{C}(\mathrm{A}, \alpha \mathrm{A}) \geq \frac{1}{2} \log \left( 1 + \mathrm{A}^2 \frac{e^{2\alpha \mu^*}}{2\pi e \sigma^2} \left( \frac{1 - e^{-\mu^*}}{\mu^*} \right)^2 \right), \tag{26}$$

*and upper-bounded by each of the two bounds*

$$\mathrm{C}(\mathrm{A}, \alpha \mathrm{A}) \le \frac{1}{2} \log\left(1 + \alpha(1-\alpha)\frac{\mathrm{A}^2}{\sigma^2}\right), \tag{27}$$

$$\mathrm{C}(\mathrm{A}, \alpha \mathrm{A}) \le \left(1 - \mathcal{Q}\left(\frac{\delta + \alpha \mathrm{A}}{\sigma}\right) - \mathcal{Q}\left(\frac{\delta + (1-\alpha)\mathrm{A}}{\sigma}\right)\right)$$
$$\cdot \log\left(\frac{\mathrm{A}}{\sigma} \cdot \frac{e^{\frac{\mu\delta}{\mathrm{A}}} - e^{-\mu\left(1+\frac{\delta}{\mathrm{A}}\right)}}{\sqrt{2\pi}\mu\left(1 - 2\mathcal{Q}\left(\frac{\delta}{\sigma}\right)\right)}\right)$$
$$- \frac{1}{2} + \mathcal{Q}\left(\frac{\delta}{\sigma}\right) + \frac{\delta}{\sqrt{2\pi}\sigma}e^{-\frac{\delta^2}{2\sigma^2}} + \frac{\sigma}{\mathrm{A}}\frac{\mu}{\sqrt{2\pi}}\left(e^{-\frac{\delta^2}{2\sigma^2}} - e^{-\frac{(\mathrm{A}+\delta)^2}{2\sigma^2}}\right)$$
$$+ \mu\alpha\left(1 - 2\mathcal{Q}\left(\frac{\delta + \frac{\mathrm{A}}{2}}{\sigma}\right)\right). \tag{28}$$

*Here, $\mu > 0$ and $\delta > 0$ are free parameters, and $\mu^*$ is the unique solution to*

$$\alpha = \frac{1}{\mu^*} - \frac{e^{-\mu^*}}{1 - e^{-\mu^*}}. \tag{29}$$

The existence and uniqueness of a solution to (29) is guaranteed by the following lemma.

**Lemma 10.** *Let $\varphi$ be a function from the positive reals to the open interval $\left(0, \frac{1}{2}\right)$*

$$\varphi : \mu \mapsto \frac{1}{\mu} - \frac{e^{-\mu}}{1 - e^{-\mu}}. \tag{30}$$

*Then, $\varphi$ is monotonically strictly decreasing and bijective with the following limiting behavior:*

$$\lim_{\mu \uparrow \infty} \varphi(\mu) = 0, \tag{31}$$

$$\lim_{\mu \downarrow 0} \varphi(\mu) = \frac{1}{2}, \tag{32}$$

$$\lim_{\mu \uparrow \infty} (\mu \cdot \varphi(\mu)) = 1. \tag{33}$$

*Proof.* See Appendix E. □

A suboptimal but useful choice for the free parameters in upper bound (28) is

$$\delta = \sigma \log\left(1 + \frac{\mathrm{A}}{\sigma}\right), \tag{34}$$

$$\mu = \mu^*\left(1 - e^{-\alpha\frac{\delta^2}{2\sigma^2}}\right), \tag{35}$$

where $\mu^*$ is the solution to (29).

Figures 1 and 2 depict the bounds of Theorem 9 for $\alpha = 0.1$ and $0.4$, where (28) is numerically minimized over $\delta, \mu > 0$.

**Theorem 11.** *If $\alpha$ lies in $\left(0, \frac{1}{2}\right)$, then*

$$\lim_{\mathrm{A} \uparrow \infty} \left\{\mathrm{C}(\mathrm{A}, \alpha \mathrm{A}) - \log\frac{\mathrm{A}}{\sigma}\right\} = -\frac{1}{2}\log 2\pi e - (1-\alpha)\mu^* - \log(1 - \alpha\mu^*) \tag{36}$$

*and*

$$\lim_{\mathrm{A} \downarrow 0} \frac{\mathrm{C}(\mathrm{A}, \alpha \mathrm{A})}{\mathrm{A}^2/\sigma^2} = \frac{\alpha(1-\alpha)}{2}. \tag{37}$$

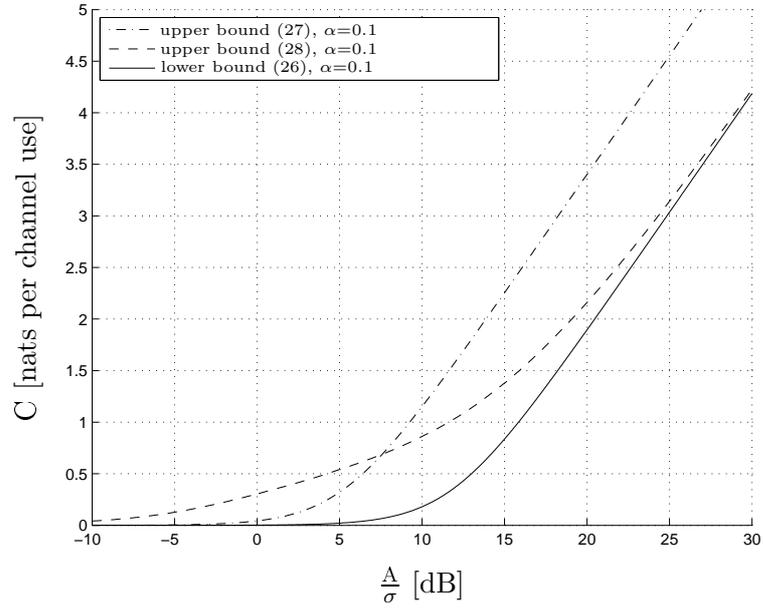

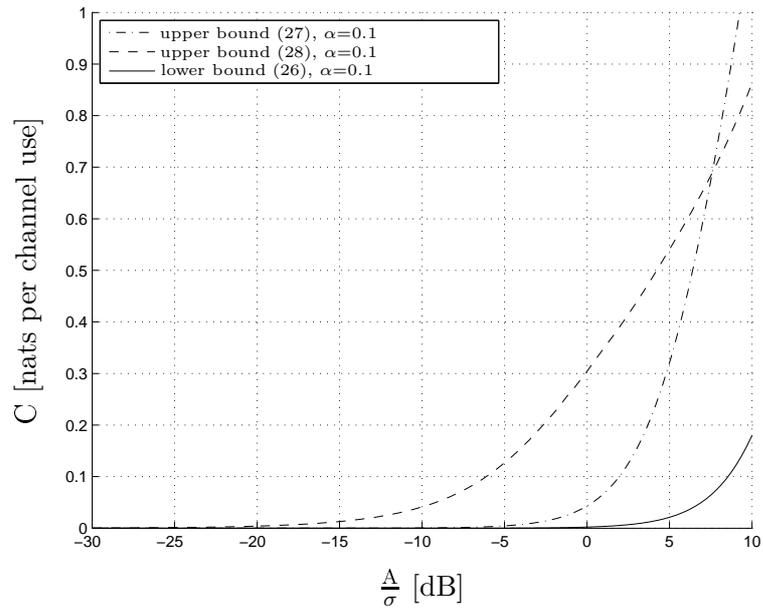

Figure 1: Bounds of Theorem 9 for $\alpha = 0.1$ when upper bound (28) is numerically minimized over $\delta, \mu > 0$. The maximum gap between upper and lower bound is 0.68 nats (for $\frac{A}{\sigma} \approx 10.5$ dB).

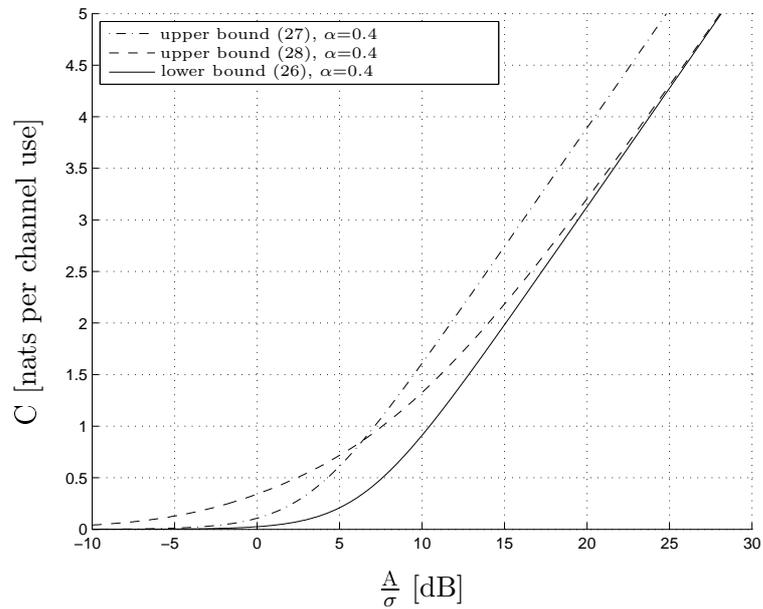

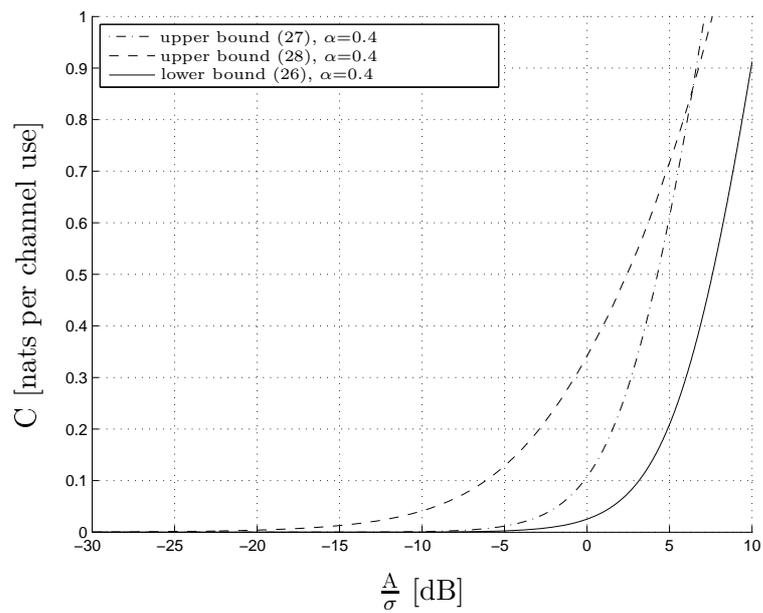

Figure 2: Bounds of Theorem 9 for $\alpha = 0.4$ with numerically optimized upper bound (28). The maximum gap between upper and lower bound is 0.52 nats (for $\frac{A}{\sigma} \approx 6.4$ dB).

## 4.2 Bounds on Channel Capacity for Case II

**Theorem 12.** *If $\alpha \in \left[\frac{1}{2}, 1\right]$, then $C(A, \alpha A)$ is lower-bounded by*

$$C(A, \alpha A) \geq \frac{1}{2} \log\left(1 + \frac{A^2}{2\pi e \sigma^2}\right), \tag{38}$$

*and is upper-bounded by each of the two bounds*

$$C(A, \alpha A) \leq \frac{1}{2} \log\left(1 + \frac{A^2}{4\sigma^2}\right), \tag{39}$$

$$C(A, \alpha A) \leq \left(1 - 2\mathcal{Q}\left(\frac{\delta + \frac{A}{2}}{\sigma}\right)\right) \log \frac{A + 2\delta}{\sigma\sqrt{2\pi}\left(1 - 2\mathcal{Q}\left(\frac{\delta}{\sigma}\right)\right)}$$
$$- \frac{1}{2} + \mathcal{Q}\left(\frac{\delta}{\sigma}\right) + \frac{\delta}{\sqrt{2\pi}\sigma} e^{-\frac{\delta^2}{2\sigma^2}}, \tag{40}$$

*where $\delta > 0$ is a free parameter.*

A useful but suboptimal choice for $\delta$ is

$$\delta = \sigma \log\left(1 + \frac{A}{\sigma}\right). \tag{41}$$

Figure 3 depicts the bounds of Theorem 12, where upper bound (40) is numerically minimized over $\delta > 0$.

**Theorem 13.** *If $\alpha$ lies in $\left[\frac{1}{2}, 1\right]$, then*

$$\lim_{A \uparrow \infty} \left\{ C(A, \alpha A) - \log \frac{A}{\sigma} \right\} = -\frac{1}{2} \log 2\pi e \tag{42}$$

*and*

$$\lim_{A \downarrow 0} \frac{C(A, \alpha A)}{A^2/\sigma^2} = \frac{1}{8}. \tag{43}$$

Note that (42) and (43) exhibit the well-known asymptotic behavior of the capacity of a Gaussian channel under a peak-power constraint only [22].

Based on the right-hand sides of (36) and (42) we define

$$\chi(\alpha) \triangleq \begin{cases} -\frac{1}{2} \log 2\pi e - (1-\alpha)\mu^* - \log(1 - \alpha\mu^*), & 0 < \alpha < \frac{1}{2}, \\ -\frac{1}{2} \log 2\pi e, & \frac{1}{2} \leq \alpha \leq 1. \end{cases} \tag{44}$$

Thus for $\alpha \in (0, 1)$, $\chi(\alpha)$ represents the second term in the high SNR asymptotic expansion of the channel capacity $C(A, \alpha A)$. It is depicted in Figure 4. Note that when $\alpha$ tends to 0, then $\chi(\alpha)$ tends to $-\infty$. This can be seen by rewriting $\chi(\alpha)$ for $\alpha \in \left(0, \frac{1}{2}\right)$ using (29) as

$$\chi(\alpha) = -\frac{1}{2} \log 2\pi e - \alpha\mu^* - \log \frac{\mu^*}{1 - e^{-\mu^*}}, \qquad \alpha \in \left(0, \frac{1}{2}\right), \tag{45}$$

and then noting that by Lemma 10 (in particular by (31) and (33)) $\mu^* \uparrow \infty$ and $\alpha\mu^* \uparrow 1$ when $\alpha \downarrow 0$.

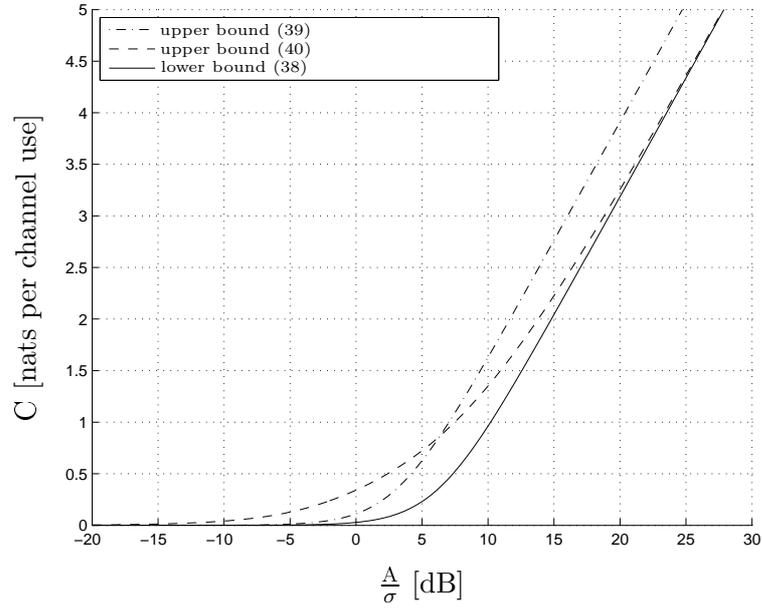

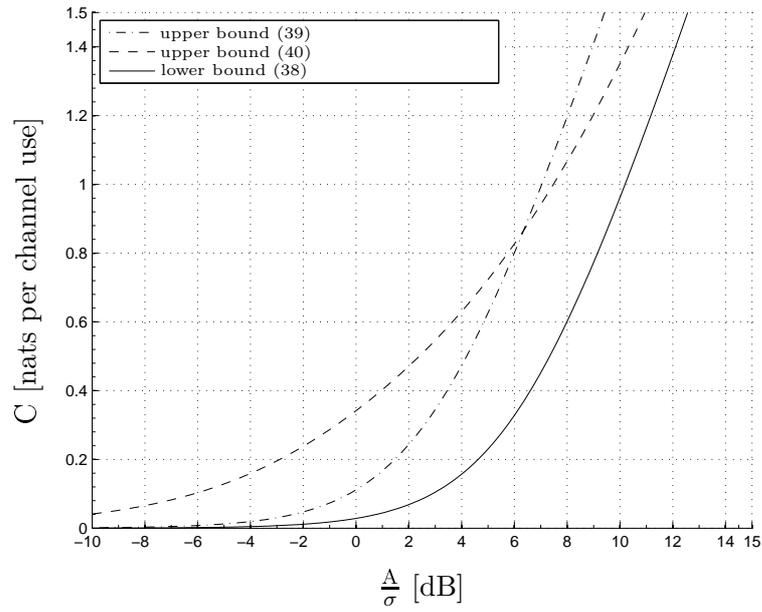

Figure 3: Bounds on capacity for $\alpha \in \left[\frac{1}{2}, 1\right]$ according to Theorem 12, where upper bound (40) is numerically minimized over $\delta > 0$. The maximum gap between upper and lower bound is 0.50 nats (for $\frac{A}{\sigma} \approx 6.4$ dB).

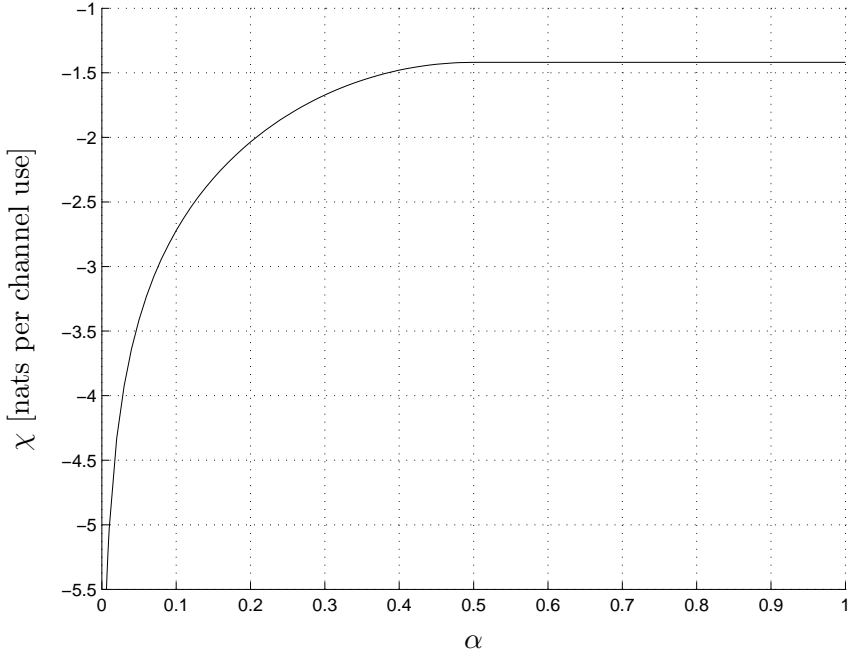

Figure 4: The term $\chi(\alpha)$ for $\alpha \in (0,1]$.

### 4.3 Bounds on Channel Capacity for Case III

**Theorem 14.** *In the absence of a peak-power constraint the channel capacity* $\mathrm{C}(\mathcal{E})$ *is lower-bounded by*

$$\mathrm{C}(\mathcal{E}) \geq \frac{1}{2} \log\left(1 + \frac{\mathcal{E}^2 e}{2\pi\sigma^2}\right), \tag{46}$$

*and is upper-bounded by each of the bounds*

$$\mathrm{C}(\mathcal{E}) \leq \log\left(\beta e^{-\frac{\delta^2}{2\sigma^2}} + \sqrt{2\pi}\sigma \mathcal{Q}\left(\frac{\delta}{\sigma}\right)\right) - \log\left(\sqrt{2\pi}\sigma\right) - \frac{\delta \mathcal{E}}{2\sigma^2}$$
$$+ \frac{\delta^2}{2\sigma^2}\left(1 - \mathcal{Q}\left(\frac{\delta}{\sigma}\right) - \frac{\mathcal{E}}{\delta}\mathcal{Q}\left(\frac{\delta}{\sigma}\right)\right) + \frac{1}{\beta}\left(\mathcal{E} + \frac{\sigma}{\sqrt{2\pi}}\right), \qquad \delta \leq -\frac{\sigma}{\sqrt{e}}, \tag{47}$$

$$\mathrm{C}(\mathcal{E}) \leq \log\left(\beta e^{-\frac{\delta^2}{2\sigma^2}} + \sqrt{2\pi}\sigma \mathcal{Q}\left(\frac{\delta}{\sigma}\right)\right) + \frac{1}{2}\mathcal{Q}\left(\frac{\delta}{\sigma}\right) + \frac{\delta}{2\sqrt{2\pi}\sigma}e^{-\frac{\delta^2}{2\sigma^2}}$$
$$+ \frac{\delta^2}{2\sigma^2}\left(1 - \mathcal{Q}\left(\frac{\delta + \mathcal{E}}{\sigma}\right)\right) + \frac{1}{\beta}\left(\delta + \mathcal{E} + \frac{\sigma}{\sqrt{2\pi}}e^{-\frac{\delta^2}{2\sigma^2}}\right)$$
$$- \frac{1}{2}\log 2\pi e\sigma^2, \qquad \delta \geq 0, \tag{48}$$

*where* $\beta > 0$ *and* $\delta$ *are free parameters. Bound (47) only holds for* $\delta \leq -\sigma e^{-\frac{1}{2}}$, *while bound (48) only holds for* $\delta \geq 0$.

A suboptimal but useful choice for the free parameters in bound (47) is

$$\delta = -2\sigma\sqrt{\log\frac{\sigma}{\mathcal{E}}}, \qquad \text{for } \frac{\mathcal{E}}{\sigma} \leq e^{-\frac{1}{4e}} \approx -0.4 \text{ dB}, \tag{49}$$

$$\beta = \frac{1}{2}\left(\mathcal{E} + \frac{\sigma}{\sqrt{2\pi}}\right) + \frac{1}{2}\sqrt{\left(\mathcal{E} + \frac{\sigma}{\sqrt{2\pi}}\right)^2 + 4\left(\mathcal{E} + \frac{\sigma}{\sqrt{2\pi}}\right)\sqrt{2\pi}\sigma e^{\frac{\delta^2}{2\sigma^2}}\mathcal{Q}\left(\frac{\delta}{\sigma}\right)}, \quad (50)$$

and for the free parameters in bound (48) is

$$\delta = \sigma \log\left(1 + \frac{\mathcal{E}}{\sigma}\right), \quad (51)$$

$$\beta = \frac{1}{2}\left(\delta + \mathcal{E} + \frac{\sigma}{\sqrt{2\pi}}e^{-\frac{\delta^2}{2\sigma^2}}\right)$$
$$+ \frac{1}{2}\sqrt{\left(\delta + \mathcal{E} + \frac{\sigma}{\sqrt{2\pi}}e^{-\frac{\delta^2}{2\sigma^2}}\right)^2 + 4\left(\delta + \mathcal{E} + \frac{\sigma}{\sqrt{2\pi}}e^{-\frac{\delta^2}{2\sigma^2}}\right)\sqrt{2\pi}\sigma e^{\frac{\delta^2}{2\sigma^2}}\mathcal{Q}\left(\frac{\delta}{\sigma}\right)}. \quad (52)$$

Figure 5 depicts the bounds of Theorem 14 when the upper bounds (47) and (48) are numerically minimized over the allowed values of $\beta$ and $\delta$.

**Theorem 15.** *In the case of only an average-power constraint,*

$$\lim_{\mathcal{E}\uparrow\infty}\left\{\mathrm{C}(\mathcal{E}) - \log\frac{\mathcal{E}}{\sigma}\right\} = \frac{1}{2}\log\frac{e}{2\pi} \quad (53)$$

*and*

$$\varlimsup_{\mathcal{E}\downarrow 0}\frac{\mathrm{C}(\mathcal{E})}{\frac{\mathcal{E}}{\sigma}\sqrt{\log\frac{\sigma}{\mathcal{E}}}} \leq 2, \quad (54)$$

$$\varliminf_{\mathcal{E}\downarrow 0}\frac{\mathrm{C}(\mathcal{E})}{\frac{\mathcal{E}}{\sigma}\sqrt{\log\frac{\sigma}{\mathcal{E}}}} \geq \frac{1}{\sqrt{2}}. \quad (55)$$

Note that the asymptotic upper and lower bound at low SNR do not coincide in the sense that their ratio equals $2\sqrt{2}$ instead of 1.

## 5 Derivation of the Firm Lower Bounds

To derive the lower bounds in Section 4 we use the entropy power inequality.

**Lemma 16 (Entropy Power Inequality).** *If $X$ and $Z$ are independent random variables with densities, then*

$$e^{2h(X+Z)} \geq e^{2h(X)} + e^{2h(Z)}. \quad (56)$$

*Proof.* See [18, Theorem 17.7.3]. □

One can find a lower bound on capacity by dropping the maximization and choosing an arbitrary input distribution $Q$ in (13). However, in order to get a tight bound, this choice of $Q$ should yield a mutual information that is reasonably close to capacity. Such a choice is difficult to find and might make the evaluation of $I(Q, W)$ intractable, because already for relatively "easy" distributions $Q$ the corresponding mutual information is difficult to compute. We circumvent these problems by using

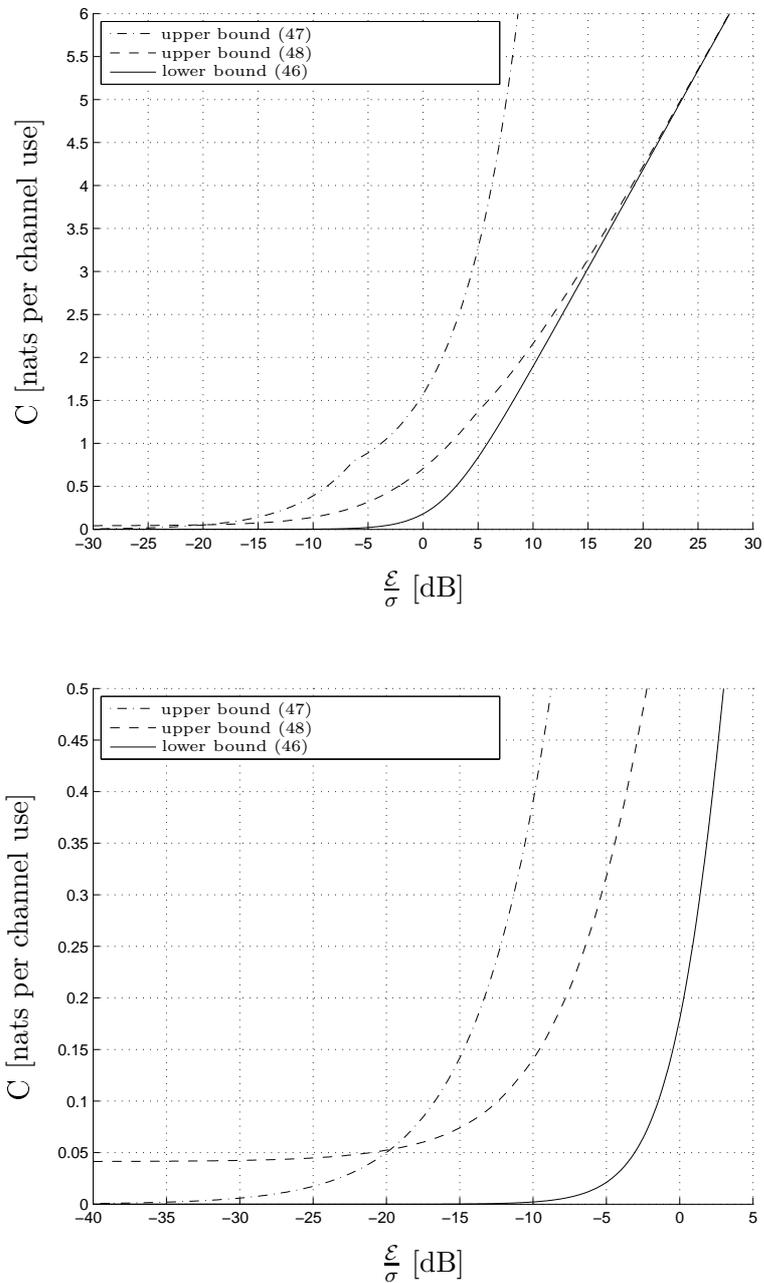

Figure 5: Bounds on capacity according to Theorem 14 when upper bounds (47) and (48) are numerically minimized over allowed values of $\beta, \delta$. The maximum gap between upper and lower bound is 0.57 nats (for $\frac{\mathcal{E}}{\sigma} \approx 2.8$ dB).

the entropy power inequality (56). For any probability distribution $Q$ with density

$$\mathsf{C} \geq I(Q, W) \tag{57}$$
$$= h(Y) - h(Y|X) \tag{58}$$
$$= h(X + Z) - h(Z) \tag{59}$$
$$\geq \frac{1}{2} \log \left( e^{2h(X)} + e^{2h(Z)} \right) - h(Z) \tag{60}$$
$$= \frac{1}{2} \log \left( 1 + \frac{e^{2h(X)}}{2\pi e \sigma^2} \right), \tag{61}$$

where (60) follows from Lemma 16. To make this lower bound as tight as possible we will choose a distribution $Q$ that maximizes differential entropy under the given constraints [18, Ch. 12].

## 5.1 Lower Bound (26) of Theorem 9

The input distribution $Q_1$ that maximizes differential entropy under a nonnegativity constraint, a peak constraint, and an average constraint has the density [18]

$$\frac{1}{\mathsf{A}} \cdot \frac{\mu^*}{1 - e^{-\mu^*}} e^{-\frac{\mu^* x}{\mathsf{A}}}, \qquad 0 \leq x \leq \mathsf{A}, \tag{62}$$

where $\mu^*$ has to be chosen such that the average-power constraint is satisfied, *i.e.*, $\mu^*$ is given as the solution to

$$\alpha = \frac{1}{\mu^*} - \frac{e^{-\mu^*}}{1 - e^{-\mu^*}}. \tag{63}$$

By Lemma 10 such a solution always exists for $0 < \alpha < \frac{1}{2}$ and is unique. The bound (26) now follows from (61) by computing $h(X)$ under the probability law $Q_1$.

## 5.2 Lower Bound (38) of Theorem 12

The uniform distribution over $[0, \mathsf{A}]$ maximizes differential entropy under a nonnegativity and a peak constraint [18]. We choose $Q_2$ to be this uniform distribution and note that then $\mathsf{E}_{Q_2}[X] = \frac{\mathsf{A}}{2}$, and hence $Q_2$ satisfies any average-power constraint larger than $\frac{\mathsf{A}}{2}$. Lower bound (38) follows directly from (61) by computing $h(X)$ under the law $Q_2$.

Notice that the uniform distribution $Q_2$ represents the limit of the input distribution $Q_1$ in Section 5.1 when $\alpha \uparrow \frac{1}{2}$. Indeed, by Lemma 10, and in particular by the limit (32),

$$\lim_{\alpha \uparrow \frac{1}{2}} \mu^* = 0, \tag{64}$$

and hence when $\alpha \uparrow \frac{1}{2}$ the density of $Q_1$ converges pointwise to the uniform density over $[0, \mathsf{A}]$.

## 5.3 Lower Bound (46) of Theorem 14

The distribution $Q_3$ that maximizes differential entropy under a nonnegativity constraint and an expectation constraint is the exponential density [18]

$$\frac{1}{\mathcal{E}} e^{-\frac{x}{\mathcal{E}}}, \qquad x \geq 0. \tag{65}$$

The bound (46) follows from (61) by computing $h(X)$ under the law $Q_3$.

Note that the density of $Q_3$ represents the pointwise limit of the density of $Q_1$ when $\alpha \downarrow 0$. Indeed, by Lemma 10, and in particular by (31) and (33),

$$\lim_{\alpha \downarrow 0} \mu^* = \infty, \tag{66}$$

$$\lim_{\alpha \downarrow 0} (\alpha \mu^*) = 1. \tag{67}$$

Also, using $\alpha = \frac{\mathcal{E}}{\mathsf{A}}$ we can rewrite the density of $Q_1$, i.e., (62), as

$$\frac{\alpha \mu^*}{\mathcal{E}} \cdot \frac{1}{1 - e^{-\mu^*}} e^{-\frac{\alpha \mu^*}{\mathcal{E}} x}, \qquad 0 \leq x \leq \frac{\mathcal{E}}{\alpha}, \tag{68}$$

which by (66) and (67) tends to (65) when $\alpha \downarrow 0$.

## 6 Derivation of the Firm Upper Bounds

The derivation of the upper bounds in Section 4 is based on Proposition 1:

$$\mathsf{C} \leq \sup_Q \mathsf{E}_Q \big[ D\big(W(\cdot|X) \| R(\cdot)\big) \big] \tag{69}$$

$$= \sup_Q \mathsf{E}_Q \left[ -\int_{-\infty}^{\infty} \log \, \mathrm{d}R(y) \, \mathrm{d}W(y|X) \right] - \frac{1}{2} \log 2\pi e \sigma^2. \tag{70}$$

Hence, we need to specify a distribution $R$, evaluate the integral (70), and finally upper-bound the supremum in (70). To upper-bound the supremum we shall present upper bounds on $\mathsf{E}_Q\left[ -\int_{-\infty}^{\infty} \log \, \mathrm{d}R(y) \, \mathrm{d}W(y|X) \right]$ which hold for arbitrary input laws $Q$ satisfying the imposed power constraints.

### 6.1 Upper Bound (27) of Theorem 9

To derive the first upper bound (27) we choose an output distribution $R_1$ corresponding to a Gaussian random variable of mean $\mathcal{E}$ and of variance $(\sigma^2 + (\mathsf{A} - \mathcal{E})\mathcal{E})$, i.e., $R_1$ has density:

$$f_1(y) = \frac{1}{\sqrt{2\pi \left(\sigma^2 + \mathcal{E}(\mathsf{A} - \mathcal{E})\right)}} e^{-\frac{(y-\mathcal{E})^2}{2\sigma^2 + 2\mathcal{E}(\mathsf{A}-\mathcal{E})}}, \qquad y \in \mathbb{R}. \tag{71}$$

For arbitrary law $Q$ satisfying $\mathsf{E}_Q[X] \leq \mathcal{E}$ and $Q(X > \mathsf{A}) = 0$ this yields

$$\mathsf{E}_Q\left[-\int_{-\infty}^{\infty} \log f_1(y)\, dW(y|X)\right]$$
$$= \log\sqrt{2\pi(\sigma^2 + \mathcal{E}(\mathsf{A}-\mathcal{E}))} + \mathsf{E}_Q\left[\frac{X^2 + \sigma^2 - 2\mathcal{E}X + \mathcal{E}^2}{2\sigma^2 + 2\mathcal{E}(\mathsf{A}-\mathcal{E})}\right] \tag{72}$$
$$\leq \log\sqrt{2\pi(\sigma^2 + \mathcal{E}(\mathsf{A}-\mathcal{E}))} + \mathsf{E}_Q\left[\frac{(\mathsf{A}-2\mathcal{E})X + \sigma^2 + \mathcal{E}^2}{2\sigma^2 + 2\mathcal{E}(\mathsf{A}-\mathcal{E})}\right] \tag{73}$$
$$\leq \log\sqrt{2\pi(\sigma^2 + \mathcal{E}(\mathsf{A}-\mathcal{E}))} + \frac{(\mathsf{A}-2\mathcal{E})\mathcal{E} + \sigma^2 + \mathcal{E}^2}{2\sigma^2 + 2\mathcal{E}(\mathsf{A}-\mathcal{E})} \tag{74}$$
$$= \frac{1}{2}\log 2\pi e\left(\sigma^2 + \mathcal{E}(\mathsf{A}-\mathcal{E})\right), \tag{75}$$

where the first inequality follows from $X^2 \leq \mathsf{A}X$ due to the peak-power constraint, and where the second inequality follows from the average-power constraint using that $\frac{\mathcal{E}}{\mathsf{A}} = \alpha \leq \frac{1}{2}$, i.e., $\mathsf{A} - 2\mathcal{E} \geq 0$. Since the resulting upper bound in (75) does not depend on the input law $Q$, (75) also upper-bounds the supremum in (70), and (27) is proved.

### 6.2 Upper Bound (28) of Theorem 9

To derive (28) we choose the law on the output $R_2$ to have density:

$$f_2(y) = \begin{cases} \frac{1}{\sqrt{2\pi}\sigma}e^{-\frac{y^2}{2\sigma^2}}, & y < -\delta, \\ \frac{1}{\mathsf{A}} \cdot \frac{\mu\left(1-2\mathcal{Q}\left(\frac{\delta}{\sigma}\right)\right)}{e^{\frac{\mu\delta}{\mathsf{A}}} - e^{-\mu\left(1+\frac{\delta}{\mathsf{A}}\right)}} e^{-\frac{\mu y}{\mathsf{A}}}, & -\delta \leq y \leq \mathsf{A}+\delta, \\ \frac{1}{\sqrt{2\pi}\sigma}e^{-\frac{(y-\mathsf{A})^2}{2\sigma^2}}, & y > \mathsf{A}+\delta, \end{cases} \tag{76}$$

where $\delta > 0$ and $\mu > 0$ are free parameters. This leads to the following expression

$$\mathsf{E}_Q\left[-\int_{-\infty}^{\infty} \log R_2'(y)\, dW(y|X)\right]$$
$$= \underbrace{\mathsf{E}_Q\left[\int_{-\infty}^{-\delta}\frac{1}{\sqrt{2\pi}\sigma}e^{-\frac{(y-X)^2}{2\sigma^2}}\log\left(\sqrt{2\pi}\sigma e^{\frac{y^2}{2\sigma^2}}\right)dy\right]}_{c_1}$$
$$- \underbrace{\mathsf{E}_Q\left[\int_{-\delta}^{\mathsf{A}+\delta}\frac{1}{\sqrt{2\pi}\sigma}e^{-\frac{(y-X)^2}{2\sigma^2}}\log\left(\frac{e^{-\frac{\mu y}{\mathsf{A}}}}{\mathsf{A}}\frac{\mu\left(1-2\mathcal{Q}\left(\frac{\delta}{\sigma}\right)\right)}{e^{\frac{\mu\delta}{\mathsf{A}}} - e^{-\mu\left(1+\frac{\delta}{\mathsf{A}}\right)}}\right)dy\right]}_{c_2}$$
$$+ \underbrace{\mathsf{E}_Q\left[\int_{\mathsf{A}+\delta}^{\infty}\frac{1}{\sqrt{2\pi}\sigma}e^{-\frac{(y-X)^2}{2\sigma^2}}\log\left(\sqrt{2\pi}\sigma e^{\frac{(y-\mathsf{A})^2}{2\sigma^2}}\right)dy\right]}_{c_3}. \tag{77}$$

We investigate each term individually. We start with $c_1$:

$$c_1 = \mathsf{E}_Q\left[\int_{-\infty}^{-\delta}\frac{1}{\sqrt{2\pi}\sigma}e^{-\frac{(y-X)^2}{2\sigma^2}}\log\left(\sqrt{2\pi}\sigma e^{\frac{y^2}{2\sigma^2}}\right)dy\right] \tag{78}$$

$$= \mathsf{E}_Q\left[\log\left(\sqrt{2\pi}\sigma\right) \cdot \mathcal{Q}\left(\frac{\delta+X}{\sigma}\right) + \frac{1}{2\sigma^2}\int_{-\infty}^{-\delta} y^2 \frac{1}{\sqrt{2\pi}\sigma} e^{-\frac{(y-X)^2}{2\sigma^2}}\,\mathrm{d}y\right] \quad (79)$$

$$\leq \mathsf{E}_Q\left[\log\left(\sqrt{2\pi}\sigma\right) \cdot \mathcal{Q}\left(\frac{\delta+X}{\sigma}\right)\right] + \frac{1}{2\sigma^2}\int_{-\infty}^{-\delta} y^2 \frac{1}{\sqrt{2\pi}\sigma} e^{-\frac{y^2}{2\sigma^2}}\,\mathrm{d}y \quad (80)$$

$$= \mathsf{E}_Q\left[\log\left(\sqrt{2\pi}\sigma\right) \cdot \mathcal{Q}\left(\frac{\delta+X}{\sigma}\right)\right] + \frac{1}{2}\mathcal{Q}\left(\frac{\delta}{\sigma}\right) + \frac{\delta}{2\sqrt{2\pi}\sigma} e^{-\frac{\delta^2}{2\sigma^2}}, \quad (81)$$

where the inequality follows from the assumption $\delta > 0$ that ensures that $(y-x)^2 \geq y^2$ for all $x \geq 0$ and $y \leq -\delta$. Similarly we get for $c_3$:

$$c_3 = \mathsf{E}_Q\left[\int_{A+\delta}^{\infty} \frac{1}{\sqrt{2\pi}\sigma} e^{-\frac{(y-X)^2}{2\sigma^2}} \log\left(\sqrt{2\pi}\sigma e^{\frac{(y-A)^2}{2\sigma^2}}\right) \mathrm{d}y\right] \quad (82)$$

$$= \mathsf{E}_Q\left[\log\left(\sqrt{2\pi}\sigma\right) \cdot \mathcal{Q}\left(\frac{\delta+A-X}{\sigma}\right)\right]$$

$$+ \mathsf{E}_Q\left[\frac{1}{2\sigma^2}\int_{A+\delta}^{\infty} (y-A)^2 \frac{1}{\sqrt{2\pi}\sigma} e^{-\frac{(y-X)^2}{2\sigma^2}}\,\mathrm{d}y\right] \quad (83)$$

$$\leq \mathsf{E}_Q\left[\log\left(\sqrt{2\pi}\sigma\right) \cdot \mathcal{Q}\left(\frac{\delta+A-X}{\sigma}\right)\right]$$

$$+ \frac{1}{2\sigma^2}\int_{A+\delta}^{\infty} (y-A)^2 \frac{1}{\sqrt{2\pi}\sigma} e^{-\frac{(y-A)^2}{2\sigma^2}}\,\mathrm{d}y \quad (84)$$

$$= \mathsf{E}_Q\left[\log\left(\sqrt{2\pi}\sigma\right) \cdot \mathcal{Q}\left(\frac{\delta+A-X}{\sigma}\right)\right] + \frac{1}{2}\mathcal{Q}\left(\frac{\delta}{\sigma}\right) + \frac{\delta}{2\sqrt{2\pi}\sigma} e^{-\frac{\delta^2}{2\sigma^2}}. \quad (85)$$

Here, the inequality follows because $(y-x)^2 \geq (y-A)^2$ for all $x \leq A$ and $y \geq A+\delta$. Finally, for $c_2$ we have

$$c_2 = \mathsf{E}_Q\left[\int_{-\delta}^{A+\delta} \frac{1}{\sqrt{2\pi}\sigma} e^{-\frac{(y-X)^2}{2\sigma^2}} \log\left(\frac{e^{\frac{\mu\delta}{A}} - e^{-\mu\left(1+\frac{\delta}{A}\right)}}{1 - 2\mathcal{Q}\left(\frac{\delta}{\sigma}\right)} \frac{A}{\mu} e^{\frac{\mu y}{A}}\right) \mathrm{d}y\right] \quad (86)$$

$$= \mathsf{E}_Q\left[\left(1 - \mathcal{Q}\left(\frac{\delta+X}{\sigma}\right) - \mathcal{Q}\left(\frac{\delta+A-X}{\sigma}\right)\right) \log\left(\frac{e^{\frac{\mu\delta}{A}} - e^{-\mu\left(1+\frac{\delta}{A}\right)}}{1 - 2\mathcal{Q}\left(\frac{\delta}{\sigma}\right)} \frac{A}{\mu}\right)\right]$$

$$+ \mathsf{E}_Q\left[\left(1 - \mathcal{Q}\left(\frac{\delta+X}{\sigma}\right) - \mathcal{Q}\left(\frac{\delta+A-X}{\sigma}\right)\right) \frac{\mu}{A} X\right]$$

$$+ \mathsf{E}_Q\left[\frac{\mu\sigma}{A\sqrt{2\pi}} \left(e^{-\frac{(\delta+X)^2}{2\sigma^2}} - e^{-\frac{(A+\delta-X)^2}{2\sigma^2}}\right)\right]. \quad (87)$$

Plugging $c_1$, $c_2$, and $c_3$ into (77) and combining this with (70) we get the following bound:

$$\mathsf{C} \leq \mathcal{Q}\left(\frac{\delta}{\sigma}\right) + \frac{\delta}{\sqrt{2\pi}\sigma} e^{-\frac{\delta^2}{2\sigma^2}} - \frac{1}{2}$$

$$+ \mathsf{E}_Q\left[\left(1 - \mathcal{Q}\left(\frac{\delta+X}{\sigma}\right) - \mathcal{Q}\left(\frac{\delta+A-X}{\sigma}\right)\right) \log\left(\frac{A\left(e^{\frac{\mu\delta}{A}} - e^{-\mu\left(1+\frac{\delta}{A}\right)}\right)}{\sqrt{2\pi}\sigma\mu\left(1 - 2\mathcal{Q}\left(\frac{\delta}{\sigma}\right)\right)}\right)\right]$$

$$+ \mathsf{E}_Q\left[\left(1 - \mathcal{Q}\left(\frac{\delta+X}{\sigma}\right) - \mathcal{Q}\left(\frac{\delta+A-X}{\sigma}\right)\right) \frac{\mu}{A} X\right]$$

$$+ \mathsf{E}_Q\left[\frac{\mu\sigma}{A\sqrt{2\pi}} \left(e^{-\frac{(\delta+X)^2}{2\sigma^2}} - e^{-\frac{(A+\delta-X)^2}{2\sigma^2}}\right)\right]. \quad (88)$$

It is shown in Appendix F that

$$\log \left( \frac{A \left( e^{\frac{\mu \delta}{A}} - e^{-\mu\left(1+\frac{\delta}{A}\right)} \right)}{\sqrt{2\pi}\sigma\mu \left(1 - 2\mathcal{Q}\left(\frac{\delta}{\sigma}\right)\right)} \right) \geq 0 \tag{89}$$

for any values of $A, \sigma, \delta, \mu > 0$. Therefore we may use Jensen's inequality combined with the concavity shown in Lemma 4 to conclude that

$$\mathsf{E}_Q\left[ \left(1 - \mathcal{Q}\left(\frac{\delta + X}{\sigma}\right) - \mathcal{Q}\left(\frac{\delta + A - X}{\sigma}\right)\right) \log \left( \frac{A\left(e^{\frac{\mu\delta}{A}} - e^{-\mu\left(1+\frac{\delta}{A}\right)}\right)}{\sqrt{2\pi}\sigma\mu\left(1 - 2\mathcal{Q}\left(\frac{\delta}{\sigma}\right)\right)} \right) \right]$$
$$\leq \left(1 - \mathcal{Q}\left(\frac{\delta + \mathsf{E}_Q[X]}{\sigma}\right) - \mathcal{Q}\left(\frac{\delta + A - \mathsf{E}_Q[X]}{\sigma}\right)\right) \log \left( \frac{A\left(e^{\frac{\mu\delta}{A}} - e^{-\mu\left(1+\frac{\delta}{A}\right)}\right)}{\sqrt{2\pi}\sigma\mu\left(1 - 2\mathcal{Q}\left(\frac{\delta}{\sigma}\right)\right)} \right) \tag{90}$$
$$\leq \left(1 - \mathcal{Q}\left(\frac{\delta + \alpha A}{\sigma}\right) - \mathcal{Q}\left(\frac{\delta + (1-\alpha)A}{\sigma}\right)\right) \log \left( \frac{A\left(e^{\frac{\mu\delta}{A}} - e^{-\mu\left(1+\frac{\delta}{A}\right)}\right)}{\sqrt{2\pi}\sigma\mu\left(1 - 2\mathcal{Q}\left(\frac{\delta}{\sigma}\right)\right)} \right). \tag{91}$$

Here, the last inequality follows from the average-power constraint, the assumption that $\alpha \leq \frac{1}{2}$, and from the fact shown in Lemma 4 that

$$\xi \mapsto \left(1 - \mathcal{Q}\left(\frac{\delta + \xi}{\sigma}\right) - \mathcal{Q}\left(\frac{\delta + A - \xi}{\sigma}\right)\right)$$

is monotonically increasing for $0 \leq \xi \leq \frac{A}{2}$.

Next, we once more use Lemma 4 to upper-bound $\left(1 - \mathcal{Q}\left(\frac{\delta+x}{\sigma}\right) - \mathcal{Q}\left(\frac{\delta+A-x}{\sigma}\right)\right)$ by its maximum value that is taken on for $x = \frac{A}{2}$:

$$\mathsf{E}_Q\left[ \left(1 - \mathcal{Q}\left(\frac{\delta + X}{\sigma}\right) - \mathcal{Q}\left(\frac{\delta + A - X}{\sigma}\right)\right) \frac{\mu}{A} X \right]$$
$$\leq \mathsf{E}_Q\left[ \left(1 - \mathcal{Q}\left(\frac{\delta + \frac{A}{2}}{\sigma}\right) - \mathcal{Q}\left(\frac{\delta + A - \frac{A}{2}}{\sigma}\right)\right) \frac{\mu}{A} X \right] \tag{92}$$
$$\leq \left(1 - \mathcal{Q}\left(\frac{\delta + \frac{A}{2}}{\sigma}\right) - \mathcal{Q}\left(\frac{\delta + \frac{A}{2}}{\sigma}\right)\right) \frac{\mu}{A} \alpha A \tag{93}$$
$$= \mu\alpha \left(1 - 2\mathcal{Q}\left(\frac{\delta + \frac{A}{2}}{\sigma}\right)\right). \tag{94}$$

And finally, we use the monotonicity of the exponential function and the fact that $X \in [0, A]$ to show the following:

$$\mathsf{E}_Q\left[ \frac{\mu\sigma}{A\sqrt{2\pi}} \left( e^{-\frac{(\delta+X)^2}{2\sigma^2}} - e^{-\frac{(A+\delta-X)^2}{2\sigma^2}} \right) \right] \leq \frac{\mu\sigma}{A\sqrt{2\pi}} \left( e^{-\frac{\delta^2}{2\sigma^2}} - e^{-\frac{(A+\delta)^2}{2\sigma^2}} \right). \tag{95}$$

Finally, combining (88) with (91), (94), and (95) yields the bound on channel capacity given in (28).

## 6.3 Upper Bound (39) of Theorem 12

To derive bound (39) we choose a Gaussian output law $R_3$ with density

$$f_3(y) = \frac{1}{\sqrt{2\pi\left(\sigma^2 + \frac{A^2}{4}\right)}} e^{-\frac{\left(y - \frac{A}{2}\right)^2}{2\sigma^2 + \frac{A^2}{2}}}. \tag{96}$$

This yields

$$\mathsf{E}_Q\left[-\int_{-\infty}^{\infty} \log R_3'(y)\, dW(y|X)\right]$$
$$= \log\sqrt{2\pi\left(\sigma^2 + \frac{A^2}{4}\right)} + \mathsf{E}_Q\left[\frac{X^2 + \sigma^2 - AX + \frac{A^2}{4}}{2\sigma^2 + \frac{A^2}{2}}\right] \tag{97}$$
$$\leq \log\sqrt{2\pi\left(\sigma^2 + \frac{A^2}{4}\right)} + \frac{\sigma^2 + \frac{A^2}{4}}{2\sigma^2 + \frac{A^2}{2}} \tag{98}$$
$$= \frac{1}{2}\log 2\pi e\sigma^2 \left(1 + \frac{A^2}{4\sigma^2}\right), \tag{99}$$

where the inequality follows because $X^2 \leq AX$ due to the peak-power constraint. Combined with (70) this yields the claimed result.

Note that the relation $\mathsf{E}_Q[X] \leq \alpha A$ has not been used. Therefore this bound is valid for all $\alpha \in [0, 1]$ and especially for all $\alpha \in \left[\frac{1}{2}, 1\right]$.

## 6.4 Upper Bound (40) of Theorem 12

The derivation of this bound is similar to the derivation of (28). We choose an output distribution $R_4$ with density:

$$f_4(y) = \begin{cases} \frac{1}{\sqrt{2\pi}\sigma} e^{-\frac{y^2}{2\sigma^2}}, & y < -\delta, \\ \frac{1 - 2\mathcal{Q}\left(\frac{\delta}{\sigma}\right)}{A + 2\delta}, & -\delta \leq y \leq A + \delta, \\ \frac{1}{\sqrt{2\pi}\sigma} e^{-\frac{(y-A)^2}{2\sigma^2}}, & y > A + \delta, \end{cases} \tag{100}$$

where $\delta > 0$ is a free parameter. This leads to the following expression:

$$\mathsf{E}_Q\left[-\int_{-\infty}^{\infty} \log R_4'(y)\, dW(y|X)\right]$$
$$= \underbrace{\mathsf{E}_Q\left[\int_{-\infty}^{-\delta} \frac{1}{\sqrt{2\pi}\sigma} e^{-\frac{(y-X)^2}{2\sigma^2}} \log\left(\sqrt{2\pi}\sigma e^{\frac{y^2}{2\sigma^2}}\right) dy\right]}_{c_1}$$
$$+ \underbrace{\mathsf{E}_Q\left[\int_{-\delta}^{A+\delta} \frac{1}{\sqrt{2\pi}\sigma} e^{-\frac{(y-X)^2}{2\sigma^2}} \log \frac{A + 2\delta}{1 - 2\mathcal{Q}\left(\frac{\delta}{\sigma}\right)} dy\right]}_{\tilde{c}_2}$$
$$+ \underbrace{\mathsf{E}_Q\left[\int_{A+\delta}^{\infty} \frac{1}{\sqrt{2\pi}\sigma} e^{-\frac{(y-X)^2}{2\sigma^2}} \log\left(\sqrt{2\pi}\sigma e^{\frac{(y-A)^2}{2\sigma^2}}\right) dy\right]}_{c_3}. \tag{101}$$

We have already upper-bounded $c_1$ and $c_3$ in (81) and (85) (assuming that $\delta > 0$). Similarly to $c_2$, we compute $\tilde{c}_2$ as follows:

$$\tilde{c}_2 = \mathsf{E}_Q\left[\int_{-\delta}^{A+\delta} \frac{1}{\sqrt{2\pi}\sigma} e^{-\frac{(y-X)^2}{2\sigma^2}} \log \frac{A+2\delta}{1-2\mathcal{Q}\left(\frac{\delta}{\sigma}\right)} dy\right] \quad (102)$$

$$= \mathsf{E}_Q\left[\left(1 - \mathcal{Q}\left(\frac{\delta+X}{\sigma}\right) - \mathcal{Q}\left(\frac{\delta+A-X}{\sigma}\right)\right) \log \frac{A+2\delta}{1-2\mathcal{Q}\left(\frac{\delta}{\sigma}\right)}\right]. \quad (103)$$

Plugging $c_1$, $\tilde{c}_2$, and $c_3$ into (101) and combining this with (70) we get

$$\mathsf{C} \leq \mathsf{E}_Q\left[\left(1 - \mathcal{Q}\left(\frac{\delta+X}{\sigma}\right) - \mathcal{Q}\left(\frac{\delta+A-X}{\sigma}\right)\right) \log \frac{A+2\delta}{\sqrt{2\pi}\sigma\left(1-2\mathcal{Q}\left(\frac{\delta}{\sigma}\right)\right)}\right]$$
$$- \frac{1}{2} + \mathcal{Q}\left(\frac{\delta}{\sigma}\right) + \frac{\delta}{\sqrt{2\pi}\sigma} e^{-\frac{\delta^2}{2\sigma^2}}. \quad (104)$$

For any $A > 0$ and $\delta > 0$

$$\frac{A+2\delta}{\sqrt{2\pi}\sigma\left(1-2\mathcal{Q}\left(\frac{\delta}{\sigma}\right)\right)} \geq \frac{2\frac{\delta}{\sigma}}{\sqrt{2\pi}\left(1-2\mathcal{Q}\left(\frac{\delta}{\sigma}\right)\right)} \geq 1, \quad (105)$$

where the first inequality follows from dropping A and the second inequality is proven in Appendix F, Equation (273). Hence,

$$\log \frac{A+2\delta}{\sqrt{2\pi}\sigma\left(1-2\mathcal{Q}\left(\frac{\delta}{\sigma}\right)\right)} \geq 0, \quad (106)$$

and we can use Lemma 4 to upper-bound $\left(1 - \mathcal{Q}\left(\frac{\delta+x}{\sigma}\right) - \mathcal{Q}\left(\frac{\delta+A-x}{\sigma}\right)\right)$ by its maximum value that is taken on for $x = \frac{A}{2}$:

$$\mathsf{E}_Q\left[\left(1 - \mathcal{Q}\left(\frac{\delta+X}{\sigma}\right) - \mathcal{Q}\left(\frac{\delta+A-X}{\sigma}\right)\right) \log \frac{A+2\delta}{\sqrt{2\pi}\sigma\left(1-2\mathcal{Q}\left(\frac{\delta}{\sigma}\right)\right)}\right]$$
$$\leq \left(1 - \mathcal{Q}\left(\frac{\delta+\frac{A}{2}}{\sigma}\right) - \mathcal{Q}\left(\frac{\delta+A-\frac{A}{2}}{\sigma}\right)\right) \log \frac{A+2\delta}{\sqrt{2\pi}\sigma\left(1-2\mathcal{Q}\left(\frac{\delta}{\sigma}\right)\right)} \quad (107)$$

$$= \left(1 - 2\mathcal{Q}\left(\frac{\delta+\frac{A}{2}}{\sigma}\right)\right) \log \frac{A+2\delta}{\sqrt{2\pi}\sigma\left(1-2\mathcal{Q}\left(\frac{\delta}{\sigma}\right)\right)}. \quad (108)$$

Again we have not used the relation $\mathsf{E}_Q[X] \leq \alpha A$, and hence the bound is valid for arbitrary $\alpha \in [0,1]$.

### 6.5 Upper Bound (47) of Theorem 14

One of the main challenges of deriving the upper bounds of Theorem 14 using duality is that without a peak-power constraint the input can be arbitrarily large (albeit with small probability). This makes it much harder to find bounds on expressions like $\mathsf{E}_Q[X^2]$. Still, we shall derive upper bound (47) using duality.

We choose a distribution $R_5$ with density

$$f_5(y) = \begin{cases} \frac{1}{\beta e^{-\frac{\delta^2}{2\sigma^2}} + \sqrt{2\pi}\sigma\mathcal{Q}\left(\frac{\delta}{\sigma}\right)} e^{-\frac{y^2}{2\sigma^2}}, & y < -\delta, \\ \frac{1}{\beta e^{-\frac{\delta^2}{2\sigma^2}} + \sqrt{2\pi}\sigma\mathcal{Q}\left(\frac{\delta}{\sigma}\right)} e^{-\frac{\delta^2}{2\sigma^2}} e^{-\frac{y+\delta}{\beta}}, & y \geq -\delta, \end{cases} \quad (109)$$

where $\delta \in \mathbb{R}$ and $\beta > 0$ are free parameters. This leads to the following expression:

$$\mathsf{E}_Q\left[-\int_{-\infty}^{\infty} \log R_5'(y)\, \mathrm{d}W(y|X)\right]$$
$$= \log\left(\beta e^{-\frac{\delta^2}{2\sigma^2}} + \sqrt{2\pi}\sigma \mathcal{Q}\left(\frac{\delta}{\sigma}\right)\right) + \mathsf{E}_Q\left[\frac{1}{2\sigma^2}\int_{-\infty}^{-\delta} y^2 \frac{1}{\sqrt{2\pi}\sigma} e^{-\frac{(y-X)^2}{2\sigma^2}}\, \mathrm{d}y\right]$$
$$+ \mathsf{E}_Q\left[\frac{\delta^2}{2\sigma^2}\int_{-\delta}^{\infty} \frac{1}{\sqrt{2\pi}\sigma} e^{-\frac{(y-X)^2}{2\sigma^2}}\, \mathrm{d}y\right]$$
$$+ \mathsf{E}_Q\left[\frac{1}{\beta}\int_{-\delta}^{\infty} (y+\delta)\frac{1}{\sqrt{2\pi}\sigma} e^{-\frac{(y-X)^2}{2\sigma^2}}\, \mathrm{d}y\right] \tag{110}$$
$$= \log\left(\beta e^{-\frac{\delta^2}{2\sigma^2}} + \sqrt{2\pi}\sigma \mathcal{Q}\left(\frac{\delta}{\sigma}\right)\right)$$
$$+ \underbrace{\mathsf{E}_Q\left[\left(\frac{1}{2} + \frac{X^2}{2\sigma^2}\right)\mathcal{Q}\left(\frac{\delta+X}{\sigma}\right) + \frac{\delta-X}{2\sqrt{2\pi}\sigma}e^{-\frac{(\delta+X)^2}{2\sigma^2}}\right]}_{c_4}$$
$$+ \underbrace{\mathsf{E}_Q\left[\frac{\delta^2}{2\sigma^2}\left(1 - \mathcal{Q}\left(\frac{\delta+X}{\sigma}\right)\right)\right]}_{c_5}$$
$$+ \underbrace{\mathsf{E}_Q\left[\frac{\delta+X}{\beta}\left(1 - \mathcal{Q}\left(\frac{\delta+X}{\sigma}\right)\right) + \frac{\sigma}{\sqrt{2\pi}\beta}e^{-\frac{(\delta+X)^2}{2\sigma^2}}\right]}_{c_6}. \tag{111}$$

We now restrict the free parameter $\delta$ to satisfy

$$\delta \leq -\frac{\sigma}{\sqrt{e}} \tag{112}$$

and continue as follows. For arbitrary input law $Q$ such that $\mathsf{E}_Q[X] \leq \mathcal{E}$:

$$\mathsf{E}_Q\left[\frac{1}{2}\mathcal{Q}\left(\frac{\delta+X}{\sigma}\right)\right] \leq \frac{1}{2}; \tag{113}$$

$$\mathsf{E}_Q\left[\frac{X^2}{2\sigma^2}\mathcal{Q}\left(\frac{\delta+X}{\sigma}\right)\right] = \mathsf{E}_Q\left[\frac{X}{2\sigma} \cdot \frac{X}{\sigma}\mathcal{Q}\left(\frac{X}{\sigma} - \frac{-\delta}{\sigma}\right)\right] \tag{114}$$

$$\leq \mathsf{E}_Q\left[\frac{X}{2\sigma} \cdot \frac{-\delta}{\sigma}\right] \tag{115}$$

$$\leq -\frac{\delta\mathcal{E}}{2\sigma^2}; \tag{116}$$

$$\mathsf{E}_Q\left[\frac{\delta-X}{2\sqrt{2\pi}\sigma}e^{-\frac{(\delta+X)^2}{2\sigma^2}}\right] \leq 0; \tag{117}$$

$$\mathsf{E}_Q\left[\frac{\delta^2}{2\sigma^2}\left(1 - \mathcal{Q}\left(\frac{\delta+X}{\sigma}\right)\right)\right] = \mathsf{E}_Q\left[\frac{\delta^2}{2\sigma^2}\left(1 - \mathcal{Q}\left(\frac{X}{\sigma} - \frac{-\delta}{\sigma}\right)\right)\right] \tag{118}$$

$$\leq \mathsf{E}_Q\left[\frac{\delta^2}{2\sigma^2}\left(1 - \mathcal{Q}\left(\frac{\delta}{\sigma}\right) + \frac{X}{-\delta}\mathcal{Q}\left(\frac{\delta}{\sigma}\right)\right)\right] \tag{119}$$

$$\leq \frac{\delta^2}{2\sigma^2}\left(1 - \mathcal{Q}\left(\frac{\delta}{\sigma}\right) - \frac{\mathcal{E}}{\delta}\mathcal{Q}\left(\frac{\delta}{\sigma}\right)\right); \tag{120}$$

$$\mathsf{E}_Q\left[\frac{\delta}{\beta}\left(1 - \mathcal{Q}\left(\frac{\delta+X}{\sigma}\right)\right)\right] \leq 0; \tag{121}$$

$$\mathsf{E}_Q\left[\frac{X}{\beta}\left(1 - \mathcal{Q}\left(\frac{\delta + X}{\sigma}\right)\right)\right] \leq \mathsf{E}_Q\left[\frac{X}{\beta}\right] \leq \frac{\mathcal{E}}{\beta};\tag{122}$$

$$\mathsf{E}_Q\left[\frac{\sigma}{\sqrt{2\pi}\beta}e^{-\frac{(\delta+X)^2}{2\sigma^2}}\right] \leq \frac{\sigma}{\sqrt{2\pi}\beta}.\tag{123}$$

Here, the first inequality (113) follows from $\mathcal{Q}(\xi) \leq 1$ for all $\xi \in \mathbb{R}$; (115) follows from Lemma 5 using assumption (112); in the subsequent inequality (116) we use the average-power constraint together with (112); (117) follows from $X \geq 0$ and $\delta < 0$ (by (112)); in (119) we use Lemma 6, and the subsequent inequality (120) follows again from the power constraint together with (112); (121) is due to (112); (122) follows because $\mathcal{Q}(\xi) \geq 0$ for all $\xi \in \mathbb{R}$; and in the last inequality (123) we upper-bound $e^{-\frac{(\delta+X)^2}{2\sigma^2}}$ by 1.

Combining (113)–(123) with (111) and (70) yields the claimed result.

### 6.6 Upper Bound (48) of Theorem 14

The bound (48) follows from the same choice (109) as we have used for the bound (47). However, here we will restrict the free parameter $\delta$ to be nonnegative:

$$\delta \geq 0.\tag{124}$$

We can then bound $c_4$ as

$$c_4 = \mathsf{E}_Q\left[\frac{1}{2\sigma^2}\int_{-\infty}^{-\delta} y^2 \frac{1}{\sqrt{2\pi}\sigma} e^{-\frac{(y-X)^2}{2\sigma^2}}\,\mathrm{d}y\right]\tag{125}$$

$$\leq \mathsf{E}_Q\left[\frac{1}{2\sigma^2}\int_{-\infty}^{-\delta} y^2 \frac{1}{\sqrt{2\pi}\sigma} e^{-\frac{y^2}{2\sigma^2}}\,\mathrm{d}y\right]\tag{126}$$

$$= \frac{1}{2}\mathcal{Q}\left(\frac{\delta}{\sigma}\right) + \frac{\delta}{2\sqrt{2\pi}\sigma}e^{-\frac{\delta^2}{2\sigma^2}},\tag{127}$$

where the inequality follows from the assumption $\delta \geq 0$ and the nonnegativity of $X$ that ensure that $(\delta + X)^2 \geq \delta^2$.

Moreover, using the concavity and monotonicity of $\xi \mapsto (1 - \mathcal{Q}(\xi))$ for $\xi \geq 0$ (see Lemma 3) and Jensen's inequality, we bound $c_5$ as follows:

$$c_5 = \mathsf{E}_Q\left[\frac{\delta^2}{2\sigma^2}\left(1 - \mathcal{Q}\left(\frac{\delta+X}{\sigma}\right)\right)\right] \leq \frac{\delta^2}{2\sigma^2}\left(1 - \mathcal{Q}\left(\frac{\delta+\mathcal{E}}{\sigma}\right)\right),\tag{128}$$

and, using the nonnegativity of $\mathcal{Q}(\cdot)$ and of $X$, we get

$$c_6 = \mathsf{E}_Q\left[\frac{\delta+X}{\beta}\left(1 - \mathcal{Q}\left(\frac{\delta+X}{\sigma}\right)\right) + \frac{\sigma}{\sqrt{2\pi}\beta}e^{-\frac{(\delta+X)^2}{2\sigma^2}}\right]\tag{129}$$

$$\leq \mathsf{E}_Q\left[\frac{\delta+X}{\beta} + \frac{\sigma}{\sqrt{2\pi}\beta}e^{-\frac{\delta^2}{2\sigma^2}}\right]\tag{130}$$

$$\leq \frac{\delta+\mathcal{E}}{\beta} + \frac{\sigma}{\sqrt{2\pi}\beta}e^{-\frac{\delta^2}{2\sigma^2}}.\tag{131}$$

Combining (111), (127), (128), and (131) with (70) yields the claimed result.

# 7 Derivation of Asymptotic Results

## 7.1 High-SNR Asymptotic Expression (36) in Theorem 11

To derive (36) we choose the free parameter $\delta$ of Theorem 9 as in (34), and set the parameter $\mu$ equal to $\mu^*$, the solution to (29). Then,

$$\lim_{A\uparrow\infty} \frac{\delta}{A} = 0, \tag{132}$$

$$\lim_{A\uparrow\infty} \delta = \infty, \tag{133}$$

$$\lim_{A\uparrow\infty} \delta e^{-\frac{\delta^2}{2\sigma^2}} = 0, \tag{134}$$

and therefore from (28) follows that for $\alpha \in \left(0, \frac{1}{2}\right)$:

$$\lim_{A\uparrow\infty} \left\{ C(A, \alpha A) - \log \frac{A}{\sigma} \right\} \leq \log \frac{1 - e^{-\mu^*}}{\sqrt{2\pi}\mu^*} - \frac{1}{2} + \mu^* \alpha. \tag{135}$$

On the other hand, from (26) follows

$$\lim_{A\uparrow\infty} \left\{ C(A, \alpha A) - \log \frac{A}{\sigma} \right\} \geq -\frac{1}{2} \log 2\pi e + \alpha \mu^* + \log \frac{1 - e^{-\mu^*}}{\mu^*}. \tag{136}$$

By Equivalence (29) and basic arithmetic reformulations it can be shown that the two bounds (135) and (136) coincide and equal the limit (36).

## 7.2 High-SNR Asymptotic Expression (42) in Theorem 13

To derive (42) we use lower bound (38) and upper bound (40) for $\delta$ as in (41). By this choice of $\delta$:

$$\lim_{A\uparrow\infty} \frac{\delta}{A} = 0, \tag{137}$$

$$\lim_{A\uparrow\infty} \delta = \infty, \tag{138}$$

$$\lim_{A\uparrow\infty} \delta e^{-\frac{\delta^2}{2\sigma^2}} = 0, \tag{139}$$

$$\lim_{A\uparrow\infty} \mathcal{Q}\left(\frac{A + 2\delta}{2\sigma}\right) \log(A + 2\delta) = 0, \tag{140}$$

and hence from upper bound (40) follows

$$\lim_{A\uparrow\infty} \left\{ C(A, \alpha A) - \log \frac{A}{\sigma} \right\}$$
$$\leq \lim_{A\uparrow\infty} \left\{ \log \frac{1 + 2\frac{\delta}{A}}{1 - 2\mathcal{Q}\left(\frac{\delta}{\sigma}\right)} - 2\mathcal{Q}\left(\frac{A + 2\delta}{2\sigma}\right) \log \frac{A + 2\delta}{\sigma\sqrt{2\pi}\left(1 - \mathcal{Q}\left(\frac{\delta}{\sigma}\right)\right)} \right.$$
$$\left. - \frac{1}{2} \log 2\pi e + \mathcal{Q}\left(\frac{\delta}{\sigma}\right) + \frac{\delta}{\sqrt{2\pi}\sigma} e^{-\frac{\delta^2}{2\sigma^2}} \right\} \tag{141}$$
$$= -\frac{1}{2} \log 2\pi e. \tag{142}$$

On the other hand, by lower bound (38):

$$\lim_{A\uparrow\infty} \left\{ C(A, \alpha A) - \log \frac{A}{\sigma} \right\} \geq \lim_{A\uparrow\infty} \frac{1}{2} \log \left( \frac{\sigma^2}{A^2} + \frac{1}{2\pi e} \right) = -\frac{1}{2} \log 2\pi e. \quad (143)$$

The two bounds coincide and therefore they prove (42) in Theorem 13.

## 7.3 High-SNR Asymptotic Expression (53) in Theorem 15

To derive (53) we use bound (48) with the following choice of the free parameters $\beta$ and $\delta$:

$$\beta \triangleq \mathcal{E}, \quad (144)$$

$$\delta \triangleq \sigma \sqrt{\log \frac{\mathcal{E}}{\sigma}}, \quad (145)$$

for $\mathcal{E} \geq \sigma$. Then,

$$\lim_{\mathcal{E}\uparrow\infty} \delta = \infty, \quad (146)$$

$$\lim_{\mathcal{E}\uparrow\infty} \frac{\delta}{\mathcal{E}} = 0, \quad (147)$$

$$\lim_{\mathcal{E}\uparrow\infty} \delta e^{-\frac{\delta^2}{2\sigma^2}} = 0, \quad (148)$$

$$\lim_{\mathcal{E}\uparrow\infty} \frac{e^{\frac{\delta^2}{2\sigma^2}}}{\mathcal{E}} = 0. \quad (149)$$

Hence, we get from (48)

$$\lim_{\mathcal{E}\uparrow\infty} \left\{ C(\mathcal{E}) - \log \frac{\mathcal{E}}{\sigma} \right\} \leq \lim_{\mathcal{E}\uparrow\infty} \left\{ -\frac{\delta^2}{2\sigma^2} + \log \left( 1 + \frac{\sqrt{2\pi}\sigma e^{\frac{\delta^2}{2\sigma^2}} \mathcal{Q}\left(\frac{\delta}{\sigma}\right)}{\mathcal{E}} \right) + \frac{\delta^2}{2\sigma^2} \right.$$
$$\left. + \frac{\delta + \mathcal{E} + \frac{\sigma}{\sqrt{2\pi}} e^{-\frac{\delta^2}{2\sigma^2}}}{\mathcal{E}} - \frac{1}{2} \log 2\pi e \right\} \quad (150)$$

$$= 1 - \frac{1}{2} \log 2\pi e = \frac{1}{2} \log \frac{e}{2\pi}. \quad (151)$$

On the other hand, we get from lower bound (46) that

$$\lim_{\mathcal{E}\uparrow\infty} \left\{ C(\mathcal{E}) - \log \frac{\mathcal{E}}{\sigma} \right\} \geq \frac{1}{2} \log \frac{e}{2\pi}. \quad (152)$$

These two bounds coincide and therefore prove (53) in Theorem 15.

## 7.4 Low-SNR Asymptotic Expression (37) in Theorem 11

In order to prove the low-SNR asymptotic expression (37) in Theorem 11, we derive an asymptotic lower bound that combined with upper bound (27) yields the desired result. The lower bound we propose is based on Theorem 2 in [20]. For the channel (6) under consideration, the technical conditions A–F in [20] are fulfilled, and Theorem 2 in [20] states that for peak-constrained inputs $|X| < \mathrm{A}$ the mutual information satisfies
$$I(X;Y) = \frac{\mathsf{Var}(X)}{2\sigma^2} + o(\mathrm{A}^2), \tag{153}$$
where $o(\mathrm{A}^2)$ decreases faster to 0 than $\mathrm{A}^2$, *i.e.*,
$$\lim_{\mathrm{A}\downarrow 0} \frac{o(\mathrm{A}^2)}{\mathrm{A}^2} = 0. \tag{154}$$

We restrict attention to settings where $0 < \mathrm{A} < 1$. Then, the binary input
$$X = \begin{cases} 0, & \text{with prob. } 1 - \alpha, \\ \mathrm{A}(1-\mathrm{A}), & \text{with prob. } \alpha, \end{cases} \tag{155}$$
is nonnegative and peak-constrained, and it satisfies the average-power constraint $\mathsf{E}[X] \le \alpha \mathrm{A}$. Hence, by (153) and since for the choice of $X$ in (155) $\mathsf{Var}(X) = \alpha(1-\alpha)\mathrm{A}^2(1-\mathrm{A})^2$,
$$I(X;Y) = \frac{\alpha(1-\alpha)\mathrm{A}^2(1-\mathrm{A})^2}{2\sigma^2} + o(\mathrm{A}^2), \qquad 0 < \mathrm{A} < 1, \quad 0 < \alpha < \frac{1}{2}, \tag{156}$$
and we obtain the following asymptotic lower bound on the capacity for $\alpha \in \left(0, \frac{1}{2}\right)$:
$$\varliminf_{\mathrm{A}\downarrow 0} \frac{C(\mathrm{A}, \alpha\mathrm{A})}{\mathrm{A}^2/\sigma^2} \ge \frac{\alpha(1-\alpha)}{2}, \qquad 0 < \alpha < \frac{1}{2}. \tag{157}$$

Furthermore, by upper bound (27) and since $\log(1+\xi) \le \xi$, for $\xi \ge 0$,
$$\varlimsup_{\mathrm{A}\downarrow 0} \frac{C(\mathrm{A}, \alpha\mathrm{A})}{\mathrm{A}^2/\sigma^2} \le \frac{\alpha(1-\alpha)}{2}, \qquad 0 < \alpha < \frac{1}{2}. \tag{158}$$

The low-SNR asymptotic expression (37) is then established by the last two inequalities.

## 7.5 Low-SNR Asymptotic Expression (43) in Theorem 13

To prove the low-SNR asymptotic expression (43) we derive an asymptotic lower bound which combined with upper bound (39) yields the desired result. The lower bound we propose is again based on Theorem 2 in [20].

We choose a nonnegative and peak-limited binary input $X$ which equiprobably takes on the values 0 and $\mathrm{A}(1-\mathrm{A})$, for $0 < \mathrm{A} < 1$. Then, we apply the same steps as in the previous section, and in analogy to (156) obtain
$$I(X;Y) = \frac{\mathrm{A}^2(1-\mathrm{A})^2}{8\sigma^2} + o(\mathrm{A}^2), \qquad 0 < \mathrm{A} < 1, \tag{159}$$
and
$$\varliminf_{\mathrm{A}\downarrow 0} \frac{C(\mathrm{A}, \alpha\mathrm{A})}{\mathrm{A}^2/\sigma^2} \ge \frac{1}{8}, \qquad \frac{1}{2} \le \alpha \le 1. \tag{160}$$

Furthermore, by upper bound (39) and from $\log(1+\xi) \le \xi$, for all $\xi \ge 0$,
$$\varlimsup_{\mathrm{A}\downarrow 0} \frac{C(\mathrm{A}, \alpha\mathrm{A})}{\mathrm{A}^2/\sigma^2} \le \frac{1}{8}, \qquad \frac{1}{2} \le \alpha \le 1. \tag{161}$$

The low-SNR asymptotic expression (43) now follows by the last two inequalities.

## 7.6 Low-SNR Asymptotic Expressions (54) in Theorem 15

The asymptotic upper bound (54) follows from upper bound (47). We choose $\delta$ as in (49), *i.e.*,

$$\delta \triangleq -2\sigma\sqrt{\log \frac{\sigma}{\mathcal{E}}}, \tag{162}$$

and

$$\beta \triangleq \frac{1}{\mathcal{E}}. \tag{163}$$

Then, from (47):

$$\frac{\mathrm{C}(\mathcal{E})}{\frac{\mathcal{E}}{\sigma}\sqrt{\log \frac{\sigma}{\mathcal{E}}}} \leq \frac{\log\left(\frac{1}{\mathcal{E}\sqrt{2\pi}\sigma}e^{-\frac{\delta^2}{2\sigma^2}} + \mathcal{Q}\left(\frac{\delta}{\sigma}\right)\right)}{\frac{\mathcal{E}}{\sigma}\sqrt{\log \frac{\sigma}{\mathcal{E}}}} + \frac{2\sigma\sqrt{\log \frac{\sigma}{\mathcal{E}}} \cdot \mathcal{E}}{2\sigma^2 \cdot \frac{\mathcal{E}}{\sigma}\sqrt{\log \frac{\sigma}{\mathcal{E}}}}$$

$$+ \frac{2\log \frac{\sigma}{\mathcal{E}}}{\frac{\mathcal{E}}{\sigma}\sqrt{\log \frac{\sigma}{\mathcal{E}}}}\left(\underbrace{1 - \mathcal{Q}\left(\frac{\delta}{\sigma}\right)}_{=\mathcal{Q}\left(-\frac{\delta}{\sigma}\right)} + \frac{\mathcal{E}}{2\sigma\sqrt{\log \frac{\sigma}{\mathcal{E}}}}\mathcal{Q}\left(\frac{\delta}{\sigma}\right)\right)$$

$$+ \frac{\mathcal{E}\left(\mathcal{E} + \frac{\sigma}{\sqrt{2\pi}}\right)}{\frac{\mathcal{E}}{\sigma}\sqrt{\log \frac{\sigma}{\mathcal{E}}}} \tag{164}$$

$$= \frac{\log\left(\frac{\mathcal{E}}{\sqrt{2\pi}\sigma^3} + \mathcal{Q}\left(\frac{\delta}{\sigma}\right)\right)}{\frac{\mathcal{E}}{\sigma}\sqrt{\log \frac{\sigma}{\mathcal{E}}}} + 1 + \frac{2\sqrt{\log \frac{\sigma}{\mathcal{E}}}}{\frac{\mathcal{E}}{\sigma}}\mathcal{Q}\left(-\frac{\delta}{\sigma}\right) + \mathcal{Q}\left(\frac{\delta}{\sigma}\right)$$

$$+ \frac{\sigma\left(\mathcal{E} + \frac{\sigma}{\sqrt{2\pi}}\right)}{\sqrt{\log \frac{\sigma}{\mathcal{E}}}}. \tag{165}$$

Next we note that

$$\lim_{\mathcal{E}\downarrow 0}\sqrt{\log \frac{\sigma}{\mathcal{E}}} = \infty, \tag{166}$$

$$\lim_{\mathcal{E}\downarrow 0}\mathcal{Q}\left(\frac{\delta}{\sigma}\right) = 1, \tag{167}$$

$$\lim_{\mathcal{E}\downarrow 0}\frac{\sqrt{\log \frac{\sigma}{\mathcal{E}}}\mathcal{Q}\left(-\frac{\delta}{\sigma}\right)}{\frac{\mathcal{E}}{\sigma}} = 0, \tag{168}$$

and, using $\mathcal{Q}(\xi) \leq 1$ and $\log(1+\xi) \leq \xi$ for all $\xi \geq 0$, that

$$\lim_{\mathcal{E}\downarrow 0}\frac{\log\left(\frac{\mathcal{E}}{\sqrt{2\pi}\sigma^3} + \mathcal{Q}\left(\frac{\delta}{\sigma}\right)\right)}{\frac{\mathcal{E}}{\sigma}\sqrt{\log \frac{\sigma}{\mathcal{E}}}} \leq \lim_{\mathcal{E}\downarrow 0}\frac{\log\left(\frac{\mathcal{E}}{\sqrt{2\pi}\sigma^3} + 1\right)}{\frac{\mathcal{E}}{\sigma}\sqrt{\log \frac{\sigma}{\mathcal{E}}}} \tag{169}$$

$$\leq \lim_{\mathcal{E}\downarrow 0}\frac{\frac{\mathcal{E}}{\sqrt{2\pi}\sigma^3}}{\frac{\mathcal{E}}{\sigma}\sqrt{\log \frac{\sigma}{\mathcal{E}}}} \tag{170}$$

$$= 0. \tag{171}$$

Together with (165) this leads to

$$\lim_{\mathcal{E}\downarrow 0}\frac{\mathrm{C}(\mathcal{E})}{\frac{\mathcal{E}}{\sigma}\sqrt{\log \frac{\sigma}{\mathcal{E}}}} \leq 2. \tag{172}$$

## 7.7 Low-SNR Asymptotic Expressions (55) in Theorem 15

We shall derive a new asymptotic lower bound at low powers which proves (55). The lower bound is obtained by lower-bounding the mutual information $I(Q, W)$ for $Q$ the probability measure with probability mass function

$$q(x) = \begin{cases} 1 - \frac{\mathcal{E}}{x_1}, & \text{if } x = 0, \\ \frac{\mathcal{E}}{x_1}, & \text{if } x = x_1, \end{cases} \tag{173}$$

where for sufficiently small $\mathcal{E}$ we choose

$$x_1 \triangleq \sigma \sqrt{c \log \frac{\sigma}{\mathcal{E}}}, \tag{174}$$

for some constant $c > 2$. Note that $x_1 \uparrow \infty$ as $\mathcal{E} \downarrow 0$. In the remaining of this section we assume $\frac{\mathcal{E}}{\sigma} \leq \frac{1}{2}$ so that the probability mass function in (173) is well-defined. The probability density of the channel output $Y$ corresponding to the input with probability mass function (173) is given by

$$f_Y(y) = \left(1 - \frac{\mathcal{E}}{x_1}\right) \frac{1}{\sqrt{2\pi\sigma^2}} e^{-\frac{y^2}{2\sigma^2}} + \frac{\mathcal{E}}{x_1} \cdot \frac{1}{\sqrt{2\pi\sigma^2}} e^{-\frac{(y-x_1)^2}{2\sigma^2}}. \tag{175}$$

In order to evaluate the mutual information $I(Q, W)$ for the chosen binary input distribution we write it as (see [25])

$$I(Q, W) = \int D\big(W(\cdot|x) \big\| W(\cdot|0)\big) \, dQ(x) - D\big(R(\cdot) \big\| W(\cdot|0)\big). \tag{176}$$

We can then evaluate the first term on the right-hand side as

$$\int D\big(W(\cdot|x) \big\| W(\cdot|0)\big) \, dQ(x)$$
$$= \frac{\mathcal{E}}{x_1} D\big(W(\cdot|x_1) \big\| W(\cdot|0)\big) \tag{177}$$
$$= \frac{\mathcal{E}}{x_1} \int_{-\infty}^{\infty} \frac{1}{\sqrt{2\pi\sigma^2}} e^{-\frac{(y-x_1)^2}{2\sigma^2}} \log\left(\frac{e^{-\frac{(y-x_1)^2}{2\sigma^2}}}{e^{-\frac{y^2}{2\sigma^2}}}\right) dy \tag{178}$$
$$= \frac{\mathcal{E}}{x_1} \int_{-\infty}^{\infty} \frac{1}{\sqrt{2\pi\sigma^2}} e^{-\frac{(y-x_1)^2}{2\sigma^2}} \left(\frac{yx_1}{\sigma^2} - \frac{x_1^2}{2\sigma^2}\right) dy \tag{179}$$
$$= \frac{\mathcal{E}}{x_1} \left(\frac{x_1^2}{\sigma^2} - \frac{x_1^2}{2\sigma^2}\right) \tag{180}$$
$$= \frac{\mathcal{E} x_1}{2\sigma^2} = \frac{\sqrt{c}}{2} \cdot \frac{\mathcal{E}}{\sigma} \sqrt{\log \frac{\sigma}{\mathcal{E}}}. \tag{181}$$

Evaluating the second term is more difficult, and in fact we only derive an upper bound on it which exhibits the desired asymptotic behavior at low SNR. We shall show that

$$\varlimsup_{\mathcal{E} \downarrow 0} \frac{D\big(R(\cdot) \big\| W(\cdot|0)\big)}{\frac{\mathcal{E}}{\sigma} \sqrt{\log \frac{\sigma}{\mathcal{E}}}} \leq \frac{\sqrt{c}}{2} - \frac{1}{\sqrt{c}}, \tag{182}$$

from which follows by (176) and (181)

$$\varliminf_{\mathcal{E} \downarrow 0} \frac{I(Q, W)}{\frac{\mathcal{E}}{\sigma} \sqrt{\log \frac{\sigma}{\mathcal{E}}}} \geq \frac{1}{\sqrt{c}}. \tag{183}$$

The desired asymptotic lower bound in (55) then follows because (183) holds for any $c > 2$.

Thus, in the remaining of this section we wish to prove (182). To this end, we write

$$D\bigl(R(\cdot)\big\|W(\cdot|0)\bigr)$$

$$= \int_{-\infty}^{\infty} f_Y(y) \log \left( \frac{\left(1 - \frac{\mathcal{E}}{x_1}\right) \frac{1}{\sqrt{2\pi\sigma^2}} e^{-\frac{y^2}{2\sigma^2}} + \frac{\mathcal{E}}{x_1} \frac{1}{\sqrt{2\pi\sigma^2}} e^{-\frac{(y-x_1)^2}{2\sigma^2}}}{\frac{1}{\sqrt{2\pi\sigma^2}} e^{-\frac{y^2}{2\sigma^2}}} \right) dy \quad (184)$$

$$= \underbrace{\int_{-\infty}^{\frac{x_1}{2}} f_Y(y) \log \left(1 - \frac{\mathcal{E}}{x_1} + \frac{\mathcal{E}}{x_1} e^{\frac{yx_1}{\sigma^2} - \frac{x_1^2}{2\sigma^2}}\right) dy}_{c_7}$$

$$+ \underbrace{\int_{\frac{x_1}{2}}^{\frac{x_1}{2}+\frac{x_1}{c}} f_Y(y) \log \left(1 - \frac{\mathcal{E}}{x_1} + \frac{\mathcal{E}}{x_1} e^{\frac{yx_1}{\sigma^2} - \frac{x_1^2}{2\sigma^2}}\right) dy}_{c_8}$$

$$+ \underbrace{\int_{\frac{x_1}{2}+\frac{x_1}{c}}^{\infty} f_Y(y) \log \left(1 - \frac{\mathcal{E}}{x_1} + \frac{\mathcal{E}}{x_1} e^{\frac{yx_1}{\sigma^2} - \frac{x_1^2}{2\sigma^2}}\right) dy}_{c_9} \quad (185)$$

and upper-bound $c_7$, $c_8$, and $c_9$. We start with upper-bounding $c_7$ where $y \geq \frac{x_1}{2}$:

$$c_7 \leq \int_{-\infty}^{\frac{x_1}{2}} f_Y(y) \log \left(1 - \frac{\mathcal{E}}{x_1} + \frac{\mathcal{E}}{x_1} e^{\frac{x_1}{\sigma^2} \frac{x_1}{2} - \frac{x_1^2}{2\sigma^2}}\right) dy \quad (186)$$

$$= \int_{-\infty}^{\xi_1} f_Y(y) \log 1 \, dy = 0, \quad (187)$$

and hence,

$$\varlimsup_{\mathcal{E} \downarrow 0} \frac{c_7}{\mathcal{E} \sqrt{\log(1/\mathcal{E})}} \leq 0. \quad (188)$$

Next we examine $c_8$. Using $\frac{\mathcal{E}}{x_1} \geq 0$ and $\log(1+\xi) \leq \xi$ for all $\xi \geq 0$ we get

$$\log \left(1 - \frac{\mathcal{E}}{x_1} + \frac{\mathcal{E}}{x_1} e^{\frac{yx_1}{\sigma^2} - \frac{x_1^2}{2\sigma^2}}\right) \leq \log \left(1 + \frac{\mathcal{E}}{x_1} e^{\frac{yx_1}{\sigma^2} - \frac{x_1^2}{2\sigma^2}}\right) \leq \frac{\mathcal{E}}{x_1} e^{\frac{yx_1}{\sigma^2} - \frac{x_1^2}{2\sigma^2}}, \quad (189)$$

and hence,

$$c_8 \leq \int_{\frac{x_1}{2}}^{\frac{x_1}{2}+\frac{x_1}{c}} f_Y(y) \frac{\mathcal{E}}{x_1} e^{\frac{yx_1}{\sigma^2} - \frac{x_1^2}{2\sigma^2}} dy \quad (190)$$

$$= \int_{\frac{x_1}{2}}^{\frac{x_1}{2}+\frac{x_1}{c}} \left(\left(1 - \frac{\mathcal{E}}{x_1}\right) \frac{\mathcal{E}}{x_1} \frac{1}{\sqrt{2\pi\sigma^2}} e^{-\frac{y^2}{2\sigma^2}} \right.$$

$$\left. + \left(\frac{\mathcal{E}}{x_1}\right)^2 \frac{1}{\sqrt{2\pi\sigma^2}} e^{-\frac{(y-x_1)^2}{2\sigma^2}}\right) e^{\frac{yx_1}{\sigma^2} - \frac{x_1^2}{2\sigma^2}} dy \quad (191)$$

$$= \left(1 - \frac{\mathcal{E}}{x_1}\right) \frac{\mathcal{E}}{x_1} \int_{\frac{x_1}{2}}^{\frac{x_1}{2}+\frac{x_1}{c}} \frac{1}{\sqrt{2\pi\sigma^2}} e^{-\frac{(y-x_1)^2}{2\sigma^2}} dy$$

$$+\left(\frac{\mathcal{E}}{x_1}\right)^2 e^{\frac{x_1^2}{\sigma^2}} \int_{\frac{x_1}{2}}^{\frac{x_1}{2}+\frac{x_1}{c}} \frac{1}{\sqrt{2\pi\sigma^2}} e^{-\frac{(y-2x_1)^2}{2\sigma^2}} \, dy \tag{192}$$

$$= \left(1 - \frac{\mathcal{E}}{x_1}\right) \frac{\mathcal{E}}{x_1} \left( \mathcal{Q}\left(\frac{x_1}{2\sigma} - \frac{x_1}{c\sigma}\right) - \mathcal{Q}\left(\frac{x_1}{2\sigma}\right) \right)$$

$$+ \left(\frac{\mathcal{E}}{x_1}\right)^2 e^{\frac{x_1^2}{\sigma^2}} \left( \mathcal{Q}\left(\frac{3x_1}{2\sigma} - \frac{x_1}{c\sigma}\right) - \mathcal{Q}\left(\frac{3x_1}{2\sigma}\right) \right) \tag{193}$$

$$= \left(1 - \frac{\frac{\mathcal{E}}{\sigma}}{\sqrt{c}\sqrt{\log\frac{\sigma}{\mathcal{E}}}}\right) \frac{\frac{\mathcal{E}}{\sigma}}{\sqrt{c}\sqrt{\log\frac{\sigma}{\mathcal{E}}}} \left( \mathcal{Q}\left(\left(\frac{\sqrt{c}}{2} - \frac{1}{\sqrt{c}}\right)\sqrt{\log\frac{\sigma}{\mathcal{E}}}\right) \right.$$
$$\left. - \mathcal{Q}\left(\frac{\sqrt{c}}{2}\sqrt{\log\frac{\sigma}{\mathcal{E}}}\right) \right)$$

$$+ \frac{\left(\frac{\mathcal{E}}{\sigma}\right)^{2-c}}{c \log\frac{\sigma}{\mathcal{E}}} \left( \mathcal{Q}\left(\left(\frac{3\sqrt{c}}{2} - \frac{1}{\sqrt{c}}\right)\sqrt{\log\frac{\sigma}{\mathcal{E}}}\right) - \mathcal{Q}\left(\frac{3\sqrt{c}}{2}\sqrt{\log\frac{\sigma}{\mathcal{E}}}\right) \right) \tag{194}$$

where in the last equality we used the definition of $x_1$. We analyze the limiting behaviors of the two summands separately. For the first term

$$\lim_{\mathcal{E} \downarrow 0} \frac{\left(1 - \frac{\frac{\mathcal{E}}{\sigma}}{\sqrt{c}\sqrt{\log\frac{\sigma}{\mathcal{E}}}}\right) \frac{\frac{\mathcal{E}}{\sigma}}{\sqrt{c}\sqrt{\log\frac{\sigma}{\mathcal{E}}}} \left( \mathcal{Q}\left(\left(\frac{\sqrt{c}}{2} - \frac{1}{\sqrt{c}}\right)\sqrt{\log\frac{\sigma}{\mathcal{E}}}\right) - \mathcal{Q}\left(\frac{\sqrt{c}}{2}\sqrt{\log\frac{\sigma}{\mathcal{E}}}\right) \right)}{\frac{\mathcal{E}}{\sigma}\sqrt{\log\frac{\sigma}{\mathcal{E}}}} = 0. \tag{195}$$

To deal with the second term we further upper-bound it using (17) and the nonnegativity of $\mathcal{Q}(\cdot)$:

$$\frac{\left(\frac{\mathcal{E}}{\sigma}\right)^{2-c}}{c \log\frac{\sigma}{\mathcal{E}}} \left( \mathcal{Q}\left(\left(\frac{3\sqrt{c}}{2} - \frac{1}{\sqrt{c}}\right)\sqrt{\log\frac{\sigma}{\mathcal{E}}}\right) - \underbrace{\mathcal{Q}\left(\frac{3\sqrt{c}}{2}\sqrt{\log\frac{\sigma}{\mathcal{E}}}\right)}_{\geq 0} \right)$$

$$< \frac{\left(\frac{\mathcal{E}}{\sigma}\right)^{2-c}}{c \log\frac{\sigma}{\mathcal{E}}} \cdot \frac{1}{\sqrt{2\pi}\sqrt{\log\frac{\sigma}{\mathcal{E}}}\left(\frac{3\sqrt{c}}{2} - \frac{1}{\sqrt{c}}\right)} \left(\frac{\mathcal{E}}{\sigma}\right)^{\frac{c}{2}\left(\frac{3}{2} - \frac{1}{c}\right)^2} \tag{196}$$

$$= \frac{\left(\frac{\mathcal{E}}{\sigma}\right)^{\frac{1}{2} + \frac{1}{2}\left(\frac{c}{4} + \frac{1}{c}\right)}}{\left(\log\frac{\sigma}{\mathcal{E}}\right)^{3/2}} \cdot \frac{1}{\sqrt{2\pi}\left(\frac{3c^{3/2}}{2} - \sqrt{c}\right)}. \tag{197}$$

Note that whenever $c > 2$ then $\left(\frac{c}{4} + \frac{1}{c}\right) > 1$ and $\left(\frac{3c^{3/2}}{2} - \sqrt{c}\right) \neq 0$, and therefore

$$\overline{\lim_{\mathcal{E} \downarrow 0}} \frac{c_8}{\frac{\mathcal{E}}{\sigma}\sqrt{\log\frac{\sigma}{\mathcal{E}}}} \leq 0. \tag{198}$$

Finally, we examine the limiting behavior of $c_9$. To this end we rewrite $c_9$ as

$$c_9 = \underbrace{\int_{\frac{x_1}{2}+\frac{x_1}{c}}^{\infty} f_Y(y) \log\left(e^{\frac{yx_1}{\sigma^2} - \frac{x_1^2}{2\sigma^2}}\right) dy}_{c_{9,1}} + \underbrace{\int_{\frac{x_1}{2}+\frac{x_1}{c}}^{\infty} f_Y(y) \log\left(\frac{\mathcal{E}}{x_1}\right) dy}_{c_{9,2}}$$

$$+ \underbrace{\int_{\frac{x_1}{2}+\frac{x_1}{c}}^{\infty} f_Y(y) \log\left(1 + \frac{\left(1 - \frac{\mathcal{E}}{x_1}\right)}{\frac{\mathcal{E}}{x_1}} e^{-\frac{yx_1}{\sigma^2} + \frac{x_1^2}{2\sigma^2}}\right) dy}_{c_{9,3}}. \tag{199}$$

We separately examine $c_{9,1}$, $c_{9,2}$, and $c_{9,3}$ and start with $c_{9,1}$:

$$c_{9,1} = \int_{\frac{x_1}{2}+\frac{x_1}{c}}^{\infty} \left(\left(1 - \frac{\mathcal{E}}{x_1}\right) \frac{1}{\sqrt{2\pi\sigma^2}} e^{-\frac{y^2}{2\sigma^2}} + \frac{\mathcal{E}}{x_1} \frac{1}{\sqrt{2\pi\sigma^2}} e^{-\frac{(y-x_1)^2}{2\sigma^2}}\right)$$
$$\cdot \left(\frac{yx_1}{\sigma^2} - \frac{x_1^2}{2\sigma^2}\right) dy \quad (200)$$

$$= \left(1 - \frac{\mathcal{E}}{x_1}\right) \int_{\frac{x_1}{2}+\frac{x_1}{c}}^{\infty} \frac{1}{\sqrt{2\pi\sigma^2}} e^{-\frac{y^2}{2\sigma^2}} \left(\frac{yx_1}{\sigma^2} - \frac{x_1^2}{2\sigma^2}\right) dy$$
$$+ \frac{\mathcal{E}}{x_1} \int_{\frac{x_1}{2}+\frac{x_1}{c}}^{\infty} \frac{1}{\sqrt{2\pi\sigma^2}} e^{-\frac{(y-x_1)^2}{2\sigma^2}} \left(\frac{x_1}{\sigma^2}(y - x_1) + \frac{x_1^2}{2\sigma^2}\right) dy \quad (201)$$

$$= \underbrace{\left(1 - \frac{\mathcal{E}}{x_1}\right)}_{\leq 1} \frac{x_1}{\sigma^2} \cdot \frac{\sigma}{\sqrt{2\pi}} e^{-\frac{x_1^2\left(\frac{1}{2}+\frac{1}{c}\right)^2}{2\sigma^2}} - \underbrace{\left(1 - \frac{\mathcal{E}}{x_1}\right)}_{\geq 0 \text{ if } \frac{\mathcal{E}}{\sigma} \leq \frac{1}{2}} \frac{x_1^2}{2\sigma^2} \underbrace{\mathcal{Q}\left(\frac{x_1}{2\sigma} + \frac{x_1}{c\sigma}\right)}_{\geq 0}$$
$$+ \frac{\mathcal{E}}{x_1} \frac{x_1}{\sigma^2} \cdot \frac{\sigma}{\sqrt{2\pi}} e^{-\frac{x_1^2}{2\sigma^2}\left(\frac{1}{c}-\frac{1}{2}\right)^2} + \frac{\mathcal{E} x_1}{2\sigma^2} \underbrace{\mathcal{Q}\left(\frac{x_1}{c\sigma} - \frac{x_1}{2\sigma}\right)}_{\leq 1} \quad (202)$$

$$\leq \frac{x_1}{\sigma\sqrt{2\pi}} e^{-\frac{x_1^2}{2\sigma^2}\left(\frac{1}{2}+\frac{1}{c}\right)^2} + \frac{\mathcal{E}}{\sigma} \cdot \frac{1}{\sqrt{2\pi}} e^{-\frac{x_1^2}{2\sigma^2}\left(\frac{1}{c}-\frac{1}{2}\right)^2} + \frac{\mathcal{E} x_1}{2\sigma^2} \quad (203)$$

$$= \frac{\sqrt{c}\sqrt{\log\frac{\sigma}{\mathcal{E}}}}{\sqrt{2\pi}} \left(\frac{\mathcal{E}}{\sigma}\right)^{\frac{c}{2}\left(\frac{1}{2}+\frac{1}{c}\right)^2} + \frac{1}{\sqrt{2\pi}} \left(\frac{\mathcal{E}}{\sigma}\right)^{\frac{c}{2}\left(\frac{1}{c}-\frac{1}{2}\right)^2+1} + \frac{\mathcal{E}}{\sigma} \cdot \frac{\sqrt{c}}{2} \sqrt{\log\frac{\sigma}{\mathcal{E}}} \quad (204)$$

where in the last step we used the definition of $x_1$ (174). Again, since $c > 2$ we have $\frac{c}{4} + \frac{1}{c} > 1$, and therefore

$$\lim_{\mathcal{E}\downarrow 0} \frac{\frac{\sqrt{c}\sqrt{\log\frac{\sigma}{\mathcal{E}}}}{\sqrt{2\pi}} \left(\frac{\mathcal{E}}{\sigma}\right)^{\frac{c}{2}\left(\frac{1}{2}+\frac{1}{c}\right)^2}}{\frac{\mathcal{E}}{\sigma}\sqrt{\log\frac{\sigma}{\mathcal{E}}}} = \lim_{\mathcal{E}\downarrow 0} \frac{\sqrt{c}}{\sqrt{2\pi}} \left(\frac{\mathcal{E}}{\sigma}\right)^{\frac{1}{2}\left(\frac{c}{4}+\frac{1}{c}\right)-\frac{1}{2}} = 0. \quad (205)$$

Moreover,

$$\lim_{\mathcal{E}\downarrow 0} \frac{\frac{1}{\sqrt{2\pi}} \left(\frac{\mathcal{E}}{\sigma}\right)^{\frac{c}{2}\left(\frac{1}{c}-\frac{1}{2}\right)^2+1}}{\frac{\mathcal{E}}{\sigma}\sqrt{\log\frac{\sigma}{\mathcal{E}}}} = \lim_{\mathcal{E}\downarrow 0} \frac{1}{\sqrt{2\pi}\sqrt{\log\frac{\sigma}{\mathcal{E}}}} \left(\frac{\mathcal{E}}{\sigma}\right)^{\frac{c}{2}\left(\frac{1}{c}-\frac{1}{2}\right)^2} = 0 \quad (206)$$

and

$$\lim_{\mathcal{E}\downarrow 0} \frac{\frac{\mathcal{E}}{\sigma} \cdot \frac{\sqrt{c}}{2}\sqrt{\log\frac{\sigma}{\mathcal{E}}}}{\frac{\mathcal{E}}{\sigma}\sqrt{\log\frac{\sigma}{\mathcal{E}}}} = \frac{\sqrt{c}}{2}, \quad (207)$$

and we conclude that

$$\varlimsup_{\mathcal{E}\downarrow 0} \frac{c_{9,1}}{\frac{\mathcal{E}}{\sigma}\sqrt{\log\frac{\sigma}{\mathcal{E}}}} \leq \frac{\sqrt{c}}{2}. \quad (208)$$

Next we analyze $c_{9,2}$. Note that $\frac{\mathcal{E}}{x_1} \leq 1$, for $\frac{\mathcal{E}}{\sigma} \leq \frac{1}{2}$, and hence $\log\frac{\mathcal{E}}{x_1} \leq 0$. Therefore,

$$c_{9,2} = \left(\underbrace{\left(1 - \frac{\mathcal{E}}{x_1}\right)}_{\geq 0 \text{ if } \frac{\mathcal{E}}{\sigma} \leq \frac{1}{2}} \underbrace{\mathcal{Q}\left(\frac{x_1}{2\sigma} + \frac{x_1}{c\sigma}\right)}_{\geq 0} + \frac{\mathcal{E}}{x_1} \mathcal{Q}\left(\frac{x_1}{c\sigma} - \frac{x_1}{2\sigma}\right)\right) \log\left(\frac{\mathcal{E}}{x_1}\right) \quad (209)$$

$$\leq \frac{\mathcal{E}}{x_1} \mathcal{Q}\left(\frac{x_1}{c\sigma} - \frac{x_1}{2\sigma}\right) \log\left(\frac{\mathcal{E}}{x_1}\right) \tag{210}$$

$$= \frac{\frac{\mathcal{E}}{\sigma}}{\sqrt{c}\sqrt{\log\frac{\sigma}{\mathcal{E}}}} \mathcal{Q}\left(\left(\frac{1}{\sqrt{c}} - \frac{\sqrt{c}}{2}\right)\sqrt{\log\frac{\sigma}{\mathcal{E}}}\right) \log\left(\frac{\frac{\mathcal{E}}{\sigma}}{\sqrt{c}\sqrt{\log\frac{\sigma}{\mathcal{E}}}}\right). \tag{211}$$

Since $c > 2$ the term $\mathcal{Q}\left(\left(\frac{1}{\sqrt{c}} - \frac{\sqrt{c}}{2}\right)\sqrt{\log\frac{\sigma}{\mathcal{E}}}\right)$ tends to 1 when $\mathcal{E}$ tends to 0, and therefore

$$\varlimsup_{\mathcal{E}\downarrow 0} \frac{c_{9,2}}{\frac{\mathcal{E}}{\sigma}\sqrt{\log\frac{\sigma}{\mathcal{E}}}} = \varlimsup_{\mathcal{E}\downarrow 0} \frac{-\log\frac{\sigma}{\mathcal{E}} - \frac{1}{2}\log c - \frac{1}{2}\log\log\frac{\sigma}{\mathcal{E}}}{\sqrt{c}\log\frac{\sigma}{\mathcal{E}}} = -\frac{1}{\sqrt{c}}. \tag{212}$$

Finally, we analyze $c_{9,3}$. Using that $\frac{x_1}{\mathcal{E}} - 1 \geq 0$ if $\frac{\mathcal{E}}{\sigma} \leq \frac{1}{2}$ we lower-bound $y \geq \frac{x_1}{2} + \frac{x_1}{c}$ to get

$$c_{9,3} = \int_{\frac{x_1}{2}+\frac{x_1}{c}}^{\infty} f_Y(y) \log\left(1 + \left(\frac{x_1}{\mathcal{E}} - 1\right)e^{-\frac{yx_1}{\sigma^2} + \frac{x_1^2}{2\sigma^2}}\right) dy \tag{213}$$

$$\leq \int_{\frac{x_1}{2}+\frac{x_1}{c}}^{\infty} f_Y(y) \log\left(1 + \left(\frac{x_1}{\mathcal{E}} - 1\right)e^{-\frac{x_1^2}{c\sigma^2}}\right) dy \tag{214}$$

$$= \left(\underbrace{\left(1 - \frac{\mathcal{E}}{x_1}\right)}_{\leq 1} \mathcal{Q}\left(\frac{x_1}{2\sigma} + \frac{x_1}{c\sigma}\right) + \frac{\mathcal{E}}{x_1} \mathcal{Q}\left(\frac{x_1}{c\sigma} - \frac{x_1}{2\sigma}\right)\right)$$

$$\cdot \log\left(1 + \underbrace{\left(\frac{x_1}{\mathcal{E}} - 1\right)}_{\leq \frac{x_1}{\mathcal{E}}} e^{-\frac{x_1^2}{c\sigma^2}}\right) \tag{215}$$

$$\leq \left(\mathcal{Q}\left(\frac{x_1}{2\sigma} + \frac{x_1}{c\sigma}\right) + \frac{\mathcal{E}}{x_1}\mathcal{Q}\left(\frac{x_1}{c\sigma} - \frac{x_1}{2\sigma}\right)\right) \log\left(1 + \frac{x_1}{\mathcal{E}}e^{-\frac{x_1^2}{c\sigma^2}}\right) \tag{216}$$

$$\leq \left(\frac{1}{2}e^{-\frac{x_1^2}{2\sigma^2}\left(\frac{1}{2}+\frac{1}{c}\right)^2} + \frac{\mathcal{E}}{x_1}\cdot\frac{1}{2}e^{-\frac{x_1^2}{2\sigma^2}\left(\frac{1}{c}-\frac{1}{2}\right)^2}\right) \cdot \left(\frac{x_1}{\mathcal{E}}e^{-\frac{x_1^2}{c\sigma^2}}\right) \tag{217}$$

$$= \left(\frac{1}{2}\left(\frac{\mathcal{E}}{\sigma}\right)^{\frac{c}{2}\left(\frac{1}{2}+\frac{1}{c}\right)^2} + \frac{1}{2\sqrt{c\log\frac{\sigma}{\mathcal{E}}}}\left(\frac{\mathcal{E}}{\sigma}\right)^{\frac{c}{2}\left(\frac{1}{c}-\frac{1}{2}\right)^2+1}\right) \sqrt{c}\sqrt{\log\frac{\sigma}{\mathcal{E}}}. \tag{218}$$

Here, (217) follows by (18) and by $\log(1+\xi) \leq \xi$, for $\xi \geq 0$.

Since for $c > 2$ we have $\frac{c}{2}\left(\frac{1}{2}+\frac{1}{c}\right)^2 > 1$ and $\left(\frac{1}{c}-\frac{1}{2}\right)^2 > 0$, we obtain the following limiting behavior:

$$\varlimsup_{\mathcal{E}\downarrow 0} \frac{c_{9,3}}{\frac{\mathcal{E}}{\sigma}\sqrt{\log\frac{\sigma}{\mathcal{E}}}} \leq 0. \tag{219}$$

By (208), (212), and (219) we conclude that

$$\varlimsup_{\mathcal{E}\downarrow 0} \frac{c_9}{\frac{\mathcal{E}}{\sigma}\sqrt{\log\frac{\sigma}{\mathcal{E}}}} \leq \frac{\sqrt{c}}{2} - \frac{1}{\sqrt{c}} \tag{220}$$

and hence, as we have set out to prove, by combining (188), (198), and (220) we obtain

$$\varlimsup_{\mathcal{E}\downarrow 0} \frac{D(f_Y \| f_{Y|X=0})}{\frac{\mathcal{E}}{\sigma}\sqrt{\log\frac{\sigma}{\mathcal{E}}}} \leq \frac{\sqrt{c}}{2} - \frac{1}{\sqrt{c}}. \tag{221}$$

## A    Proof of Lemma 4

We start by showing that $f(\cdot)$ is strictly concave. Since $\xi_0, \gamma \geq 0$, the functions $\xi \mapsto \mathcal{Q}(\xi_0 + \xi)$ and $\xi \mapsto \mathcal{Q}(\xi_0 + \gamma - \xi)$ are strictly convex over $[0, \gamma]$ by Lemma 3. Then, also the function $\xi \mapsto (\mathcal{Q}(\xi_0 + \xi) + \mathcal{Q}(\xi_0 + \gamma - \xi))$ must be strictly convex over $[0, \gamma]$ and we conclude that the function $f : \xi \mapsto (1 - \mathcal{Q}(\xi_0 + \xi) - \mathcal{Q}(\xi_0 + \gamma - \xi))$ is strictly concave.

The symmetry of $f(\cdot)$ around $\xi = \frac{\gamma}{2}$ can be seen by noting that for all $\xi' \in \left[0, \frac{\gamma}{2}\right]$

$$f\left(\frac{\gamma}{2} + \xi'\right) = 1 - \mathcal{Q}\left(\xi_0 + \frac{\gamma}{2} + \xi'\right) - \mathcal{Q}\left(\xi_0 + \gamma - \frac{\gamma}{2} - \xi'\right) \tag{222}$$

$$= 1 - \mathcal{Q}\left(\xi_0 + \frac{\gamma}{2} + \xi'\right) - \mathcal{Q}\left(\xi_0 + \frac{\gamma}{2} - \xi'\right) \tag{223}$$

is identical to

$$f\left(\frac{\gamma}{2} - \xi'\right) = 1 - \mathcal{Q}\left(\xi_0 + \frac{\gamma}{2} - \xi'\right) - \mathcal{Q}\left(\xi_0 + \gamma - \frac{\gamma}{2} + \xi'\right) \tag{224}$$

$$= 1 - \mathcal{Q}\left(\xi_0 + \frac{\gamma}{2} - \xi'\right) - \mathcal{Q}\left(\xi_0 + \frac{\gamma}{2} + \xi'\right). \tag{225}$$

Finally, that $f(\cdot)$ has its maximum at $\frac{\gamma}{2}$ and is monotonically strictly increasing over $\left[0, \frac{\gamma}{2}\right]$ follows by the symmetry and the strict concavity.

## B    Proof of Lemma 5

For $\xi \leq \mu$ we have

$$\underbrace{\xi}_{\leq \mu} \underbrace{\mathcal{Q}(\xi - \mu)}_{\leq 1} \leq \mu, \tag{226}$$

because $\xi \geq 0$ and $\mathcal{Q}(\cdot)$ is nonnegative.

Let us then assume that $\xi > \mu$ and introduce a variable substitution $y = \xi - \mu$. Then for $y > 0$, we get

$$\xi \mathcal{Q}(\xi - \mu) = (y + \mu)\mathcal{Q}(y) \tag{227}$$

$$\leq (y + \mu)\frac{1}{2}e^{-\frac{y^2}{2}} \tag{228}$$

$$\leq \frac{y}{2}e^{-\frac{y^2}{2}} + \frac{\mu}{2} \tag{229}$$

$$\leq \frac{1}{2}e^{-\frac{1}{2}} + \frac{\mu}{2} \tag{230}$$

$$\leq \mu. \tag{231}$$

Here, the first inequality (228) follows from (18); in (229) we upper-bound $e^{-\frac{y^2}{2}}$ by 1; then in (230) we replace $ye^{-\frac{y^2}{2}}$ by its maximum $e^{-\frac{1}{2}}$; and the final inequality (231) holds because we have assumed that $\mu \geq e^{-\frac{1}{2}}$.

## C    Proof of Lemma 6

Define

$$f_1(\xi) \triangleq 1 - \mathcal{Q}(\xi - \mu), \tag{232}$$

$$f_2(\xi) \triangleq 1 - \mathcal{Q}(-\mu) + \frac{\xi}{\mu}\mathcal{Q}(-\mu). \tag{233}$$

We shall show that for all $\xi \geq 0$
$$f_1(\xi) \leq f_2(\xi), \tag{234}$$
where $\mu$ is a nonnegative constant. Let us start with the case $0 \leq \xi \leq \mu$. From Lemma 3 it follows that $f_1(\cdot)$ is strictly convex over $[0, \mu]$. Moreover, note that $f_1(0) = f_2(0)$ and that the slope of $f_1(\cdot)$ at $\xi = 0$ is smaller than the slope of $f_2(\cdot)$:

$$\left.\frac{\partial}{\partial \xi} f_1(\xi)\right|_{\xi=0} = \left.\frac{1}{\sqrt{2\pi}} e^{-\frac{(\xi-\mu)^2}{2}}\right|_{\xi=0} \tag{235}$$

$$= \frac{1}{\sqrt{2\pi}} e^{-\frac{\mu^2}{2}} \tag{236}$$

$$= \frac{1}{\mu} \cdot \frac{\mu}{\sqrt{2\pi}} e^{-\frac{\mu^2}{2}} \tag{237}$$

$$\leq \frac{1}{\mu} \cdot \frac{1}{\sqrt{2\pi}} e^{-\frac{1}{2}} \tag{238}$$

$$< \frac{1}{\mu} \cdot \frac{1}{2} \tag{239}$$

$$\leq \frac{1}{\mu} \bigl(1 - \mathcal{Q}(\mu)\bigr) \tag{240}$$

$$= \frac{1}{\mu} \mathcal{Q}(-\mu) \tag{241}$$

$$= \frac{\partial}{\partial \xi} f_2(\xi). \tag{242}$$

Here, (238) follows from upper-bounding $\mu e^{-\frac{\mu^2}{2}}$ by its maximum $e^{-\frac{1}{2}}$; the subsequent inequality (239) from the fact that $\frac{1}{\sqrt{2\pi e}} < \frac{1}{2}$; and the subsequent two steps (240) and (241) follow from Lemma 3.

Hence for small values of $\xi$, $f_1(\xi) \leq f_2(\xi)$. Since $f_1$ is convex, it can intersect with the linear function $f_2$ at most one more time apart from the intersection at $\xi = 0$. However, for $\xi = \mu$,

$$f_1(\mu) = \frac{1}{2}, \tag{243}$$
$$f_2(\mu) = 1, \tag{244}$$

i.e., $f_2$ is still larger than $f_1$, so no intersection has taken place in the interval $(0, \mu]$.

For $\xi > \mu$, no intersection can take place either because by definition

$$f_2(\xi) > 1 > f_1(\xi), \quad \xi > \mu. \tag{245}$$

This completes the proof.

## D  Proof of Lemma 8

We first state an auxiliary proposition which is used to prove the lemma.

**Proposition 17.** *Let the random variable $X$ take value in the interval $[0, A]$, and let $Z \sim \mathcal{N}_\mathbb{R}(0, \sigma^2)$ be independent of $X$. Then, there exists a random variable $\tilde{X}$ taking value in $[0, A]$ and independent of $Z$ that satisfies*

$$\mathsf{E}\bigl[\tilde{X}\bigr] = \frac{1}{2} A \tag{246}$$

*and*
$$I(\tilde{X}; \tilde{X} + Z) \geq I(X; X + Z). \tag{247}$$

*Proof.* Define $\bar{X} = \mathrm{A} - X$ and note that

$$\begin{align}
I(X; X + Z) &= I(X; \mathrm{A} - X - Z) \tag{248}\\
&= I(X; \mathrm{A} - X + Z) \tag{249}\\
&= I(\mathrm{A} - X; \mathrm{A} - X + Z) \tag{250}\\
&= I(\bar{X}; \bar{X} + Z), \tag{251}
\end{align}$$

where (248) and (250) follow because $I(U; V) = I(U; g(V))$ whenever $g$ is one-to-one; where (249) follows from the symmetry of the centered Gaussian; and where (251) follows from the definition of $\bar{X}$.

Let $B$ be a binary random variable that takes on the values 0 and 1 equiprobably and independently of the pair $(X, Z)$. Define the random variable $\tilde{X}$ equal to $X$ when $B = 1$ and equal to $\bar{X}$ when $B = 0$. We show that $\tilde{X}$ (which takes value in $[0, \mathrm{A}]$) satisfies both (246) and (247). Condition (246) follows by the total law of expectation, by the definition of $\tilde{X}$, by the independence of $B$ and $(X, \bar{X})$, and because $\mathsf{E}[\bar{X}] = \mathrm{A} - \mathsf{E}[X]$:

$$\mathsf{E}\left[\tilde{X}\right] = \frac{1}{2}\mathsf{E}\left[\tilde{X} \mid B = 1\right] + \frac{1}{2}\mathsf{E}\left[\tilde{X} \mid B = 0\right] = \frac{1}{2}\mathsf{E}[X] + \frac{1}{2}\mathsf{E}\left[\bar{X}\right] = \frac{1}{2}\mathrm{A}. \tag{252}$$

□

Condition (247) follows because conditioning reduces differential entropy, because $\tilde{X}$ is independent of $(X, \bar{X}, Z)$, and by (251):

$$\begin{align}
I(\tilde{X}; \tilde{X} + Z) &= h(\tilde{X} + Z) - h(Z) \tag{253}\\
&\geq h(\tilde{X} + Z | B) - h(Z) \tag{254}\\
&= \frac{1}{2} h(\tilde{X} + Z | B = 1) + \frac{1}{2} h(\tilde{X} + Z | B = 0) - h(Z) \tag{255}\\
&= \frac{1}{2}\bigl(h(X + Z) - h(Z)\bigr) + \frac{1}{2}\bigl(h(\bar{X} + Z) - h(Z)\bigr) \tag{256}\\
&= \frac{1}{2} I(X + Z; X) + \frac{1}{2} I(\bar{X} + Z; \bar{X}) \tag{257}\\
&= I(X; X + Z). \tag{258}
\end{align}$$

With the aid of Proposition 17 we now prove Lemma 8:

*Proof (Proof of Lemma 8).* Let $Q^*$ denote the capacity-achieving input distribution (which exists by Lemma 7). Then, by Proposition 17, there exists an input distribution $\tilde{Q}$ with average power

$$\mathsf{E}_{\tilde{Q}}[X] = \frac{1}{2}\mathrm{A} \tag{259}$$

such that

$$I(\tilde{Q}, W) \geq I(Q^*, W). \tag{260}$$

Whenever $\alpha \geq \frac{1}{2}$, $\tilde{Q}$ is a valid input distribution in the optimization in (13) and by (260) it achieves capacity. But by the uniqueness of the capacity-achieving input distribution (Lemma 7) the distributions $Q^*$ and $\tilde{Q}$ must coincide, and therefore by (259)

$$\mathsf{E}_{Q^*}[X] = \frac{1}{2}\mathrm{A}. \tag{261}$$

□

# E  Proof of Lemma 10

We start by proving that the function $\varphi(\cdot)$ is monotonically strictly decreasing. This follows immediately by taking the derivative of $\varphi(\cdot)$

$$\varphi'(\mu) = -\frac{1}{\mu^2} - \frac{e^{-\mu}}{(1-e^{-\mu})^2}, \tag{262}$$

which is strictly negative for $\mu \in (0, \infty)$. Then, by the strict monotonicity and because $\varphi(\cdot)$ is continuous, $\varphi(\cdot)$ is bijective.

We are left with proving the asymptotic results. Whereas the first limit (31) follows directly, the second limit (32) follows by rewriting the function $\varphi(\cdot)$ as

$$\varphi(\mu) = \frac{1-(1+\mu)e^{-\mu}}{\mu(1-e^{-\mu})}, \tag{263}$$

and then applying two times de l'Hôpital's rule. Finally, the third limit (33) is obtained by observing that

$$\mu\varphi(\mu) = 1 - \frac{\mu e^{-\mu}}{1-e^{-\mu}} \tag{264}$$

and that $\mu e^{-\mu}$ tends to zero as $\mu$ tends to infinity. This concludes the proof of the lemma.

# F  Appendix for the Proof of the Upper Bound (28)

It remains to show that

$$\log\left(\frac{A\left(e^{\frac{\mu\delta}{A}} - e^{-\mu\left(1+\frac{\delta}{A}\right)}\right)}{\sqrt{2\pi}\sigma\mu\left(1-2\mathcal{Q}\left(\frac{\delta}{\sigma}\right)\right)}\right) \geq 0, \qquad \forall\, A, \sigma, \delta, \mu > 0. \tag{265}$$

We investigate the following expression:

$$\frac{A\left(e^{\frac{\mu\delta}{A}} - e^{-\mu\left(1+\frac{\delta}{A}\right)}\right)}{\sqrt{2\pi}\sigma\mu\left(1-2\mathcal{Q}\left(\frac{\delta}{\sigma}\right)\right)} \geq \frac{A\left(e^{\frac{\mu\delta}{A}} - e^{-\mu\frac{\delta}{A}}\right)}{\sqrt{2\pi}\sigma\mu\left(1-2\mathcal{Q}\left(\frac{\delta}{\sigma}\right)\right)} \tag{266}$$

$$= \frac{\frac{A}{\mu\delta} \cdot \frac{1}{2}\left(e^{\frac{\mu\delta}{A}} - e^{-\frac{\mu\delta}{A}}\right)}{\frac{\sqrt{2\pi}\left(1-2\mathcal{Q}\left(\frac{\delta}{\sigma}\right)\right)}{2\frac{\delta}{\sigma}}} \tag{267}$$

$$= \frac{\frac{A}{\mu\delta} \cdot \sinh\left(\frac{\mu\delta}{A}\right)}{\frac{\sqrt{2\pi}\left(1-2\mathcal{Q}\left(\frac{\delta}{\sigma}\right)\right)}{2\frac{\delta}{\sigma}}}, \tag{268}$$

where the inequality follows because we drop a factor $e^{-\mu} \leq 1$. Now we note that since $\sinh(\cdot)$ is a convex function over $[0, \infty)$ for any triple $0 \leq \xi_0 < \xi_1 < \xi_2 < \infty$, $\sinh(\xi_1)$ lies below the line segment connecting $\sinh(\xi_0)$ and $\sinh(\xi_2)$, *i.e.*,

$$\sinh(\xi_1) \leq \frac{\xi_1-\xi_0}{\xi_2-\xi_0}\left(\sinh(\xi_2) - \sinh(\xi_0)\right). \tag{269}$$

By choosing $\xi_0 = 0$ and since $\sinh(0) = 0$ we conclude that

$$\frac{\sinh(\xi_1)}{\xi_1} \leq \frac{\sinh(\xi_2)}{\xi_2}, \qquad \forall\, 0 < \xi_1 < \xi_2. \tag{270}$$

Hence the function $\xi \mapsto \frac{\sinh(\xi)}{\xi}$ is monotonically increasing over $(0, \infty)$ and has its infimum in the limit $\xi \downarrow 0$, *i.e.*,

$$\frac{\sinh(\xi)}{\xi} \geq \lim_{\xi' \downarrow 0} \frac{\sinh(\xi')}{\xi'} = 1, \qquad \forall\, \xi > 0, \tag{271}$$

where the limit follows by de l'Hôpital's rule. Similarly, since the function $\xi \mapsto \frac{\sqrt{2\pi}}{2}(1 - 2\mathcal{Q}(\xi))$ is concave over $[0, \infty)$ and since $\frac{\sqrt{2\pi}}{2}(1 - 2\mathcal{Q}(0)) = 0$,

$$\frac{\frac{\sqrt{2\pi}}{2}(1 - 2\mathcal{Q}(\xi_1))}{\xi_1} \geq \frac{\frac{\sqrt{2\pi}}{2}(1 - 2\mathcal{Q}(\xi_2))}{\xi_2}, \quad \forall\, 0 < \xi_1 < \xi_2. \tag{272}$$

Hence the function $\xi \mapsto \frac{\frac{\sqrt{2\pi}}{2}(1 - 2\mathcal{Q}(\xi))}{\xi}$ is monotonically decreasing on $(0, \infty)$ and has its supremum in the limit when $\xi \downarrow 0$, *i.e.*,

$$\frac{\sqrt{2\pi}(1 - 2\mathcal{Q}(\xi))}{2\xi} \leq \lim_{\xi' \downarrow 0} \frac{\sqrt{2\pi}(1 - 2\mathcal{Q}(\xi'))}{2\xi'} = 1, \qquad \forall\, \xi > 0, \tag{273}$$

where again the limit follows by de l'Hôpital's rule.

Thus, for any $A, \sigma, \mu, \delta > 0$:

$$\log\left(\frac{A\left(e^{\frac{\mu\delta}{A}} - e^{-\mu\left(1 + \frac{\delta}{A}\right)}\right)}{\sqrt{2\pi}\sigma\mu\left(1 - 2\mathcal{Q}\left(\frac{\delta}{\sigma}\right)\right)}\right) \geq \log\left(\frac{\frac{A}{\mu\delta} \cdot \sinh\left(\frac{\mu\delta}{A}\right)}{\frac{\sqrt{2\pi}(1 - 2\mathcal{Q}(\frac{\delta}{\sigma}))}{2\frac{\delta}{\sigma}}}\right) \tag{274}$$

$$\geq \log\left(\frac{\inf_{\xi > 0}\left\{\frac{1}{\xi} \cdot \sinh(\xi)\right\}}{\sup_{\xi > 0}\left\{\frac{\sqrt{2\pi}(1 - 2\mathcal{Q}(\xi))}{2\xi}\right\}}\right) \tag{275}$$

$$= \log\frac{1}{1} = 0. \tag{276}$$

## Acknowledgments

Fruitful discussions with Ding-Jie Lin are gratefully acknowledged.